# In the search of molecular signature of sarcopenia in *C. elegans*

**Diana David-Rus***

Department of Bioinformatics and Structural Biochemistry,
Institute of Biochemistry of Romanian Academy,Splaiul independentei 296, 060031,
Bucharest 17, Romania

## Abstract

Age-related muscle decline, a condition referred to as sarcopenia and defined as loss in muscle mass and muscle strength over time, is one of the most pervasive problems of the elderly, such that significant declines in strength and mobility affects essentially every old person We have found that aging C. *elegans* body wall muscle undergoes a process remarkably reminiscent of human sarcopenia. Both have mid-life onset and are characterized by progressive loss of sarcomeres and cytoplasmic volume; both are associated with locomotory decline. To extend understanding of this fundamental problem, I surveyed expression of all known muscle related genes to describe a profile of transcriptional changes in muscle that transpires during adult life and aging. Importantly, the intersection of this dataset with that from aging flies and some human studies can suggest conserved genes that might impact the process most strongly. Hypotheses I formulate will be used to drive experiments at the bench and perhaps to focus attention for human therapies.

Keywords: statistical methods, clustering, sarcopenia, C.*elegans*, microarray data analysis

# 1. Introduction:

Sarcopenia is an "age-related" loss of muscle mass leading to muscle weakness, limited mobility, and increased susceptibility to injury (1,2). Overall changes with age that contribute to sarcopenia include declines in androgenic and growth hormone concentrations (3), declines in spontaneous physical activity, and changes in dietary intake of protein and/or energy (4). Specifically, in skeletal muscle there is a selective loss of muscle fibers (5), decline in total muscle area and reduced muscle capillarization, shortening velocity, and maximal force (6).

To begin to identify the molecular basis for the loss of muscle mass with age, investigators have measured in mammals changes in gene expression on a global scale during aging in skeletal muscle using serial analysis of gene expression (7), cDNA arrays (8, 9), and oligonucleotide-based microarrays (10, 11, 12, 13). These studies have reported changes in gene expression consistent with decreased protein synthesis, impaired oxidative defense, and decreased activity of mitochondrial proteins. They have also reported differential expression of genes involved in energy metabolism, DNA damage repair, stress response, immune/inflammatory response, RNA binding and splicing, and proteasome degradation. Although these studies have provided insight into the age-related changes in gene expression and therefore the aging process, the human studies in particular have limitations with regard to sample size, number of genes surveyed, overall smaller differences in gene expression, and pooling of samples. Importantly, investigating the molecular mechanisms underlying sarcopenia in humans with the use of microarrays is also complicated by the inherent variability in human gene expression profiles. This variability is likely due to differences in genetics, diet, environment, and habitual patterns of activity, making it more difficult to identify true age-specific alterations. In fact, investigators using the human Affymetrix microarrays to study young vs.

older males (13) found that the intragroup (*n* = 8) variability was so high that a special ratio method was needed to be developed to reduce the within-group variance (14).

Using *C. elegans* as animal model in an effort to better understand the biology of aging we put an emphasis on mid-life changes that, we consider might influence aging, and give us an insight into sarcopenia as a process. Studies in our lab and others have suggested that critical events during the mid-life of the nematode can influence the aging of the organism.  The small nematodes have less variability in gene expression profiles, and a lot of muscle-related genes (50% have human homologie), which presents clear advantages for aging and sarcopenia  microarray studies.

 In the present study, Affymetrix  GeneChips special designed for *C. elegans* (by Hoffmann-LaRoche company, from Basel, Switzerland) were used to interrogate the expression of 18,612 genes (open reading frames).

**2. Experimental design:**

In this study we used same experimental design as described in first chapter. Our experiment includes time points covering the reproductive and post-reproductive periods, with a series of consecutive mid-life time points.

In order to be able to grow the worms in a synchronous way, we used only spe-9 (hc88), which is a temperature sensitive mutation. This strain does not produce progeny at 25 C;

We cultured *spe-9*(hc88) mutants at 25.5°C in order to avoid contamination of a synchronous culture with young animals. *spe-9*(hc88) is defective in spermatogenesis and produce unfertilized oocytes when reared at high temperature.  Any "escapers" were visually identified and eliminated manually, so cultures were highly synchronous. At days 3, 6, 9, 10, 11, 12 and 15 of culture (as measured from egg deposition) animals just reach adulthood at day 3. We harvested

~ 20,000 worms per time point in three trials and used RNA from each of these for three independent hybridizations. Because of our focus on potential relevant changes at the midlife transition, we also prepared another triplicate experiment in which we harvested nematodes at days 9, 10 and 11.

Data from these middle time-points were combined with those in the more extended trials to increase the significance of findings at days 9, 10 and 11.

Therefore, in our data we have six total independent repeats for the middle life time points day 9, 10, 11 and 3 repeats for day3, 6, 12, 15.

## 3. Data Analysis-Methods

### 3.1 Outlier detection:

In order to detect the outliers we used (as outlier exclusion test) the Nalimov outlier test. For each gene per *Condition* a modified Nalimov outlier test is performed for data points representing Replicate experiments. (see, Kaiser R, Gottschalk G (1972)).

### 3.2 Scaling

In order to achieve, scaling of the data on the chip and between chips, for each chip, we calculated the median signal intensity over all probe sets. The median of this median signal intensity from all chips was calculated. Then, every chip, is scaled to this median value.

### 3.3 KNN estimation method

The data was estimated based on the values of the K nearest neighbor genes estimator, ( Tibshirani, R., Botstein, D. & Altman, R. B. (2001).

We also transformed the data using a logarithmic transformation in base 2, X= $\log_2(X)$. The reason we do this is that is preferable to work with logged intensities rather than absolute intensities since the variation of logged intensities tends to be less dependent on the magnitude of the values; taking logs, reduces the skewness of the distributions and improves variance estimation.

## 3.4 Filtering

A special method for filtering genes based on highest variance was designed. Genes were filtered on the basis of their variation across the samples. A set of 2000 genes were chosen, on the basis of their standard deviations. I analysed also other lists choosing as, for example, 2500 genes, 3000, and 5000 genes; or filtering data using ANOVA as way of filtering and then used FDR test for checking on false positive and obtained a list of 1241 genes, however after careful unsupervised analysis of all the lists mentioned, we conclude that the lists of 2000 genes obtained from the filtering based on highest variation is more suitable for answering question in regard of aging, sarcopenia and that the ANOVA method is too conservative for our purpose. Also the lists of 3000 and 5000 genes were unnecessary large from the point of biology novelty. Therefore in presenting the results of our unsupervised as well as supervised method, we will refer to the list of 2000 genes obtained based on the highest variation filtering.

## 3.5 Normalization

In order to normalize the data, I went through two steps: first, the step of what is known in statistics literature as "center mean", and obtain this way a new vector. Then the new vector is divided to its "standard deviation", meaning normalize the newly obtained vector. (See Methods Chapter 1).

## 3.6 Unsupervised method for data mining- clustering:

I've chosen to use for our data a new approach to clustering based on the physical properties of a magnetic system. The algorithm is a Monte Carlo-based method (in particular Swedsen-Wang Monte Carlo method) and uses KNN for defining the neighbors. We were able to find clusters that have not been obtained by other unsupervised clustering methods as Tree- View or K means clustering. The reason is that this method has a number of unique advantages:

1 The number of the "macroscopic" clusters is an output of the algorithm.
2 The hierarchical organization of the data is reflected in the way the clusters split or merge when a control parameter is varied.
3 The results are insensitive to the initial conditions.

Comparing this algorithm with other clustering algorithms, the drawback of methods like "Tree View" or "K means algorithms" is that they have high sensitivity to initialization and they have poor performance when the data contains overlapping clusters; The most serious problem is lack of cluster validity criteria; none of these methods provide an index that could be used to identify the most significant partitions among those obtained in entire hierarchy ( Methods Chapter 1 and Domany et. al. Physical Review '96 for more on SPC algorithm).

### 3.7 Supervised methods

For compiling various lists I used AQL language which  is a  new query language for the Acedb database system. It borrows syntax and ideas from OQL, the ODMG's proposed query language for object-oriented databases (which is supported by O2), Lorel, a language for querying semi-structured data in the Lore database system developed at Stanford, and Boulder. I also used the Gene Ontology database, (GO_term). (http://www.geneontology.org).

## 4. Results

### 4.1  Sarcopenia signature

I have expanded on previous work in *C. elegans* (see Kim et al. 2002, Kenyon et. al 2003) studies by increasing the number of samples over the midlife time window, and, most importantly, focusing our efforts on defining a molecular signature of sarcopenia rather than a general survey of gene changes with age. We also extended our previous analysis on muscle related genes from Chapter 1.

In our sarcopenia studies I used a combination of supervised and un-supervised methods. For this purpose I compiled several lists of genes muscle- related genes. The definition of sarcopenia at the phenotypic level is that is an "age-related" loss of muscle mass leading to muscle weakness and reduce mobility.  At the genotypic level we might suspect that any changes in muscle- related genes might be the reason of an  sarcopenia phenotype. In the rest of the work I will call a sarcopenia signature any major changes in muscle-related gene expression. Given the phenotypic aspect of the sarcopenia  one might expect to identify as sarcopenia signature  a down-regulated gene expression pattern. As I will show in this work, these will not be always the case.  I will call as 'positive connection with the sarcopenia phenotype' any down-regulated gene expression pattern.

   At first I compiled a list of genes considered to be expressed in muscle cells. In order to do this I  used a combination of bioinformatics tools special designed for searches in Worm Database. I used AQL language (see Methods section) which is a new query language for the Acedb database system to compile a list of 829 genes from the worm database expressed in muscle cells. Out of the 829 gene list in our raw data we identified 721 genes. We wanted to understand how many of these 721 genes considered to be expressed in cell muscle are in our list of 2000 genes with highest variation. To obtain the list of 2000 genes I used a selection method based on the highest variation of the gene expression in conjunction with a k-nearest neighbor estimator (see Methods section).  We identified 42 genes as being at the intersection between the 721 genes with the list of 2000 genes. The set of 42 muscle expressed genes are included in the list of 2000 genes, that show greatest variation during adult life, and might serve as a signature of *C. elegans* sarcopenia.  These 42 genes might be representative as genes involved in muscle if we consider as underlying hypothesis that genes that vary the most should be more involved in the biological process than genes that show a relative steady state of gene expression. We analyzed the expression of each of the 42 genes which might comprise the signature of sarcopenia over time. The list of the 42 genes with ORF annotations, CDG annotation ( or 3 letter names) as well as a concise description for each gene as was found in worm base using AQL query language can be seen in Table 1.

| 'T14A8.1' | 'ric-3' | Biological process: embryonic development ending in birth or egg hatching (IMP) growth (IMP) nematode larval development (IMP) protein targeting to membrane |
| 'H30A04.1' | 'eat-20' | '"eat-20 encodes a paralog of the C. elegans and Drosophila genes crb-1 and crumbs, expressed in pharynx, head neurons, hypodermis.<br>Biological process: hermaphrodite genitalia development |
| 'C12C8.1' | 'hsp-70' | 'hsp-70 encodes a member of the hsp70 family.' Biological process determination of adult life span |
| 'F11C3.3' | 'unc-54' | '"unc-54 encodes a muscle myosin class II heavy chain (MHC B); UNC-54 is the major myosin heavy chain expressed in C. elegans ;<br>Biological process: body morphogenesis (IMP) inositol lipid-mediated signaling (IPI) locomotion (IMP) muscle contraction (IMP) muscle thick filament assembly (IMP) oviposition (IMP) pharyngeal pumping (IPI) positive regulation of locomotion |
| 'C13B9.4' | '...' | "C13B9.4 is orthologous to the human gene CALCITONIN RECEPTOR Biological process: G-protein signaling, adenylate cyclase activating pathway (ISS) cellular calcium ion homeostasis (ISS) locomotion (IMP) positive regulation of adenylate cyclase activity |
| 'ZK721.1' | 'tag-130' | The precise role is not known;<br>detected in such tissues as body wall muscle, hypodermis, intestine, pharynx, and the gonad |
| 'W02C12.3' | 'hlh-30.' | '"W02C12.3 is orthologous to the human gene TRANSCRIPTION FACTOR BINDING TO IGHM ENHANCER 3 (TFE3; OMIM:314310), Helix loop helix transcription factor EB |
| 'ZC101.2' | 'unc-52' | biological processes:<br><br>cell adhesion (IEA) cell migration (IGI) determination of adult life span (IMP) embryonic development ending in birth or egg hatching (IMP) epidermal growth factor receptor signaling |
| | | pathway (IGI) locomotion (IMP) molting cycle, collagen and cuticulin-based cuticle (IMP) muscle development (IMP) muscle morphogenesis (IGI) nematode larval development (IMP) positive regulation of growth rate. UNC-52 is synthesized by the hypodermis and localizes to the extracellular matrix between hypodermis and muscle<br><br>And following Molecular function: calcium ion binding, structural molecule activity |
| 'C16D9.2' | 'rol-3' | biological process: collagen and cuticulin-based cuticle development, embryonic development ending in birth or egg hatching, locomotion  positive regulation of growth rate (IMP) protein amino acid phosphorylation (IEA)Cellular component integral to membrane (IEA)Molecular function ATP binding (IEA) protein kinase activity (IEA) protein serine/threonine kinase activity (IEA) protein tyrosine kinase activity (IEA) |
| 'F56D12.1' | 'alh-6' | '"alh-6 is orthologous to the human gene ALDEHYDE DEHYDROGENASE 4 FAMILY, MEMBER A1 (ALDH4A1; OMIM:606811), biological process: locomotion (IMP) metabolic process, positive regulation of growth rate (IMP) positive regulation of locomotion (IMP) proline biosynthetic process (IEA) reproduction (IMP) Cellular component mitochondrial matrix (IEA)Molecular function 1-pyrroline-5-carboxylate dehydrogenase activity (IEA) oxidoreductase activity |
| 'F40F9.10' | '...' | Has larval expression: pharynx; anal depressor muscle; body wall muscle;Adult Expression: pharynx; anal depressor muscle; body wall muscle; |
| 'H28G03.6' | 'mtm-5' | MTM-5 is expressed in adult pharynx, intestine, and body wall muscle, but has no obvious function in RNAi assays |
| 'R07B7.11' | 'gana-1' | '"R07B7.11 is orthologous to the human gene ALPHA-GALACTOSIDASE B (GALB; OMIM:104170), which when mutated leads to Schindler disease. Description: gana-1 encodes a protein with homology to both human alpha-galactosidase (alpha-GAL) and alpha-N-acetylgalactosaminidase (alpha-NAGA) enzymes; GANA-1 is expressed in body wall muscle and intestinal cells and in coelomocytes;  Biological process: carbohydrate metabolic process, glycoside catabolic process , metabolic process |
| 'T05C12.10' | 'qua-1' | Biological process: cell communication , embryonic development ending in birth or egg hatching, locomotion (IMP) molting cycle, collagen and cuticulin-based cuticle |
| | | (IMP) multicellular organismal development (IEA) nematode larval development (IMP) positive regulation of multicellular organism growth (IMP) proteolysis (IEA)<u>Molecular function:</u> peptidase activity (IEA) |
| 'F48F7.1' | 'alg-1' | 'A homolog of rde-1 that is involved in RNA interference and affects developmental timing. <u>Biological process:</u> determination of adult life span (IMP) embryonic development (IGI) embryonic development ending in birth or egg hatching (IMP) hermaphrodite genitalia development (IMP) locomotion (IMP) molting cycle, collagen and cuticulin-based cuticle (IMP) nematode larval development (IMP) positive regulation of growth rate (IMP) positive regulation of locomotion (IMP) positive regulation of multicellular organism growth (IMP) vulval development |
| 'C26C6.5' | 'dcp-66' | Biological process: embryonic development ending in birth or egg hatching (IMP) growth (IMP) hermaphrodite genitalia development (IMP) locomotion (IMP) morphogenesis of an epithelium (IMP) negative regulation of vulval development (IMP) nematode larval development (IMP) positive regulation of growth rate (IMP) positive regulation of vulval development (IMP) reproduction |
| 'F11E6.2' | 'grl-24' | is expressed in body wall muscle and intestine; No gene ontology terms have been assigned to grl-24 |
| 'T01B10.2' | 'grd-14' | Biological process: locomotion (IMP) positive regulation of multicellular organism growth (IMP) vulval development |
| 'F37H8.5' | '...' | Gamma-interferon inducible lysosomal thiol reductase |
| 'F53A9.10' | 'tnt-2' | Biological process: locomotion (IMP) positive regulation of growth rate (IMP) positive regulation of locomotion (IMP) reproduction |
| 'F42G4.3' | 'zyx-1' | '"zyx-1 encodes a zyxin homolog that physically interacts with P granule components (GLH proteins); Biological process: reproduction |
| 'C44H4.4' | '...' | Uncharacterized conserved protein |
| 'Y38F1A.6' | '...' | Biological process: metabolic process (IEA) Molecular function: metal ion binding (IEA) oxidoreductase activity |
| 'D2045.9' | '...' | Biological process: hermaphrodite genitalia development (IMP) lipopolysaccharide biosynthetic process (IEA) locomotion (IMP) morphogenesis of an epithelium (IMP) positive regulation of growth rate (IMP) reproduction |
| 'Y75B8A.7' | '...' | Biological process: growth (IMP) hermaphrodite genitalia development (IMP) nematode larval development (IMP) positive regulation of growth rate (IMP) rRNA processing (IEA) reproduction |
| 'T01C8.5' | '...' | Biological process amino acid metabolic process (IEA) |
| | | biosynthetic process (IEA) positive regulation of growth rate |
| 'F25H2.1' | 'tli-1' | Description: none available |
| 'ZK112.2' | 'ncl-1' | ncl-1 encodes a B-box zinc finger protein that may be a repressor of RNA polymerase I and III transcription; has much larger neuronal nucleoli than normal |
| 'F42A10.3' | '...' | Molecular function: methyltransferase activity |
| 'F54C9.11' | '...' | Guanine nucleotide exchange factor |
| 'F07A5.7' | 'unc-15' | The unc-15 gene encodes a paramyosin ortholog; Biological function: body morphogenesis (IMP) carbohydrate metabolic process (IEA) growth (IMP) locomotion (IMP) muscle thick filament assembly (IMP) nematode larval development (IMP) oviposition (IMP) regulation of cytoskeleton organization and biogenesis |
| 'F11A1.3' | 'daf-12' | daf-12 encodes a member of the steroid hormone receptor superfamily that affects dauer formation downstream of the TGF- and insulin signaling pathways, and affects gonad-dependent adult longevity together with DAF-16. Is homologous to human VITAMIN D RECEPTOR. Biological process: negative regulation of multicellular organism growth (IMP) positive regulation of growth rate (IMP) regulation of development, heterochronic (IMP) regulation of transcription, DNA-dependent |
| 'F38E11.2' | 'hsp-12.6' | HSP-12.6 is required in vivo for normal lifespan; hsp-12.6 encodes a small heat-shock protein; HSP-12.6 is predominantly and ubiquitously expressed in L1 larvae without any obvious induction by stressors; but, in adult hermaphrodites, at least one HSP-12 is also expressed in spermatids (and perhaps spermatocytes), as well as in some vulval cells; hsp-12.6(RNAi) animals are shorter-lived than normal; |
| 'F56H9.4' | 'gpa-9' | gpa-9 encodes a member of the G protein alpha subunit family of heterotrimeric GTPases; it is expressed in ASJ, PHB, PVQ, pharyngeal muscle, and the spermatheca |
| 'C40C9.1' | 'twk-20' | Concise Description: none available; Biological process potassium ion transport |
| 'T27E4.3' | 'hsp-16.48' | hsp-16.48 encodes a 16-kD heat shock protein (HSP) that is a member of the hsp16/hsp20/alphaB-crystallin (HSP16) family of heat shock proteins Biological process determination of adult life span. |
| 'E03D2.2' | 'nlp-9' | Concise Description: none available |
| 'C02F4.2' | 'tax-6' | tax-6 encodes an ortholog of calcineurin A Biological process: chemosensory behavior (IMP) chemotaxis (IMP) dauer larval development (IGI) hyperosmotic response |
| | | (IGI) locomotion (IMP) olfactory behavior (IGI) olfactory learning (IMP) positive regulation of growth rate (IMP) positive regulation of multicellular organism growth (IMP) thermosensory behavior (IMP) thermotaxis |
| 'C36B7.7' | 'hen-1' | '"hen-1 encodes a secretory protein that contains a low-density lipoprotein receptor class A domain. GFP reporter is expressed in pharyngeal muscles, the vulva, and weakly in a subset of neurons;Biological process: associative learning |
| 'F42A10.2' | 'nfm-1' | nfm-1 encodes a homolog of human merlin/schwannomin (NF2), which when mutated leads to neurofibromatosis. |
| 'F11E6.5' | 'elo-2' | '"The elo-2 gene encodes a palmitic acid elongase, homologous to polyunsaturated fatty acid (PUFA) elongases such as ELO-1<br>Biological process: positive regulation of growth rate |
| 'T20B3.2' | 'tni-3' | Biological process: muscle contraction (IMP) nematode larval development (IMP) oviposition (IMP) post-embryonic body morphogenesis (IMP)<br>Cellular component: sarcomere |

Legend: The blue colored names are down-regulated genes

**Table 1 List of 42 genes that might be representative for sarcopenia signature.**

The next step should be to identify the gene patterns in this list. In this sense I clustered the list of 42 genes *C. elegans* using the SPC approach to identify 8 clusters, classified (see Chapter 1) based on size (number of genes in each cluster) and stability. The hierarchical organization of the data reflected in the way clusters split or merge has a graphical representation as a tree, called a dendogram ( see fig 1). For details on the clustering algorithm see Chapter 1 as well as in E. Domany et. al, Neural Computation (1997). The clusters or nodes I obtained were annotated as G1-G8, each with a distinctive pattern. Besides classifications of clusters based on the size and stability criterion mentioned above (found in size/stability table), I attempted a classification based on patterns of gene expression identified in each such cluster. The results of this clustering analysis were compiled for an easy access in a web- based design that facilitates their analysis.

The entire informational content of the web-based clustering design is displayed graphically or in tables. I will mention below some of the links which can be found in the main web page:

- heat-map graph with all the genes normalized before being clustered.

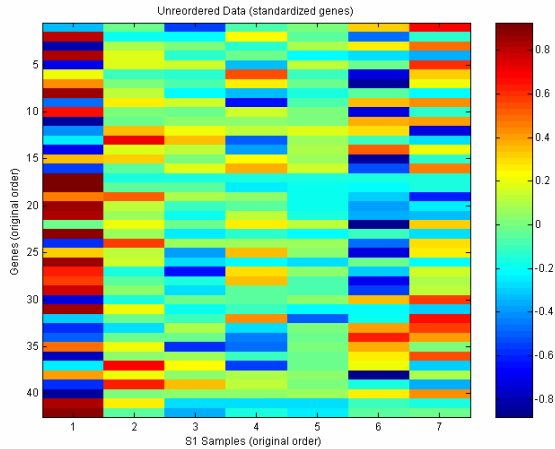

- PCA –a graph displaying the principal component analysis (see Chapter 1 for details and references on PCA method)
- Dendrogram with Stable Clusters –web based accessible dendogram
- Dendrogram next to Reordered Data (after clustering): hit-map graph & dendogram
- Reordered Genes : table with all 42 genes and the clusters where they fit.
- Samples :  time points
- Parameters for SPC
- Access to each cluster for pattern visualization and gene members.

Each cluster can be accessed from the main web page and is represented graphically in two plot formats: as a heat map and as gene expression level changes over time. In addition, a short description of the biological content, of each cluster, correspondence of the cluster with any other clusters, and the list of gene members found in the respective cluster is included. Two tables with clusters sorted based on the stability and size are also presented.

Among the 42 genes I depicted 8 clusters, using SPC clustering algorithm (see more about SPC in Chapter 1).

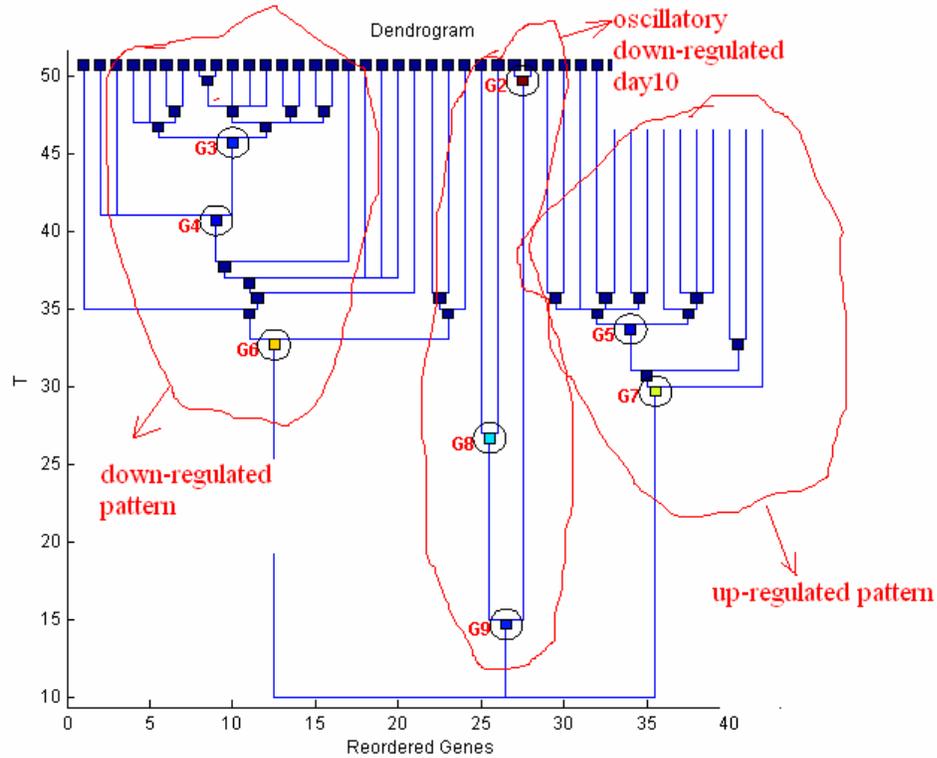

**Fig. (1). Dendogram showing 8 clusters annotated G2-G9 and the pattern categories they enter**

Fig.(1) shows  8 clusters  annotated from  G2 - G9; the dendogram shows the hierarchical organization found in the data based on SPC algorithm.

Below is the dendogram and the heat map graph of the 42 genes;

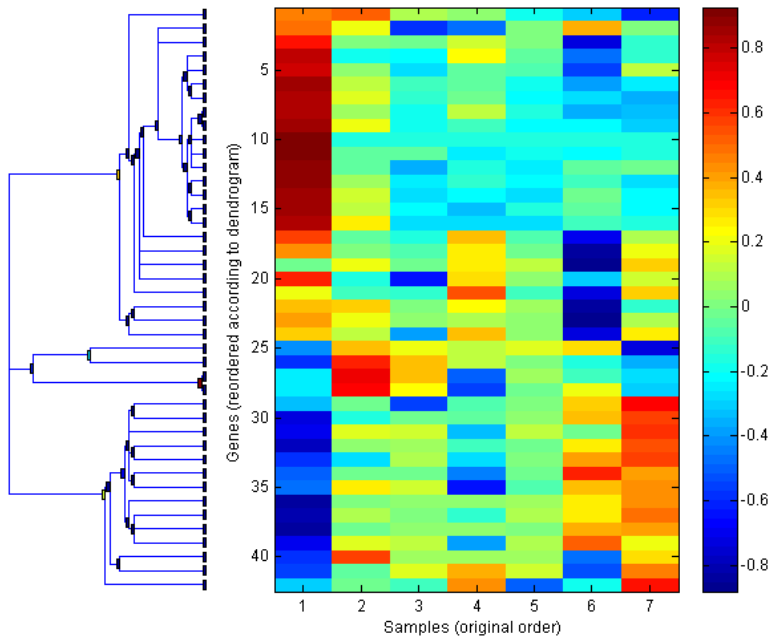

**Legend: x-axis**-7 time points: 1/day3; 2/day6; 3/day9; 4/day10; 5/day11; 6/day12; 7/day15; **y-axis**-42 gene expressions; color bar: from red-high gene expression too dark blue-low gene expressions.

**Fig. 2 Dendogram and heat map of the 42 genes after clustering.**

I depicted 3 main patterns in the clustered data: a down-regulated pattern, oscillatory pattern and up-regulated pattern. The dendogram in fig.1 shows the 8 clusters and the pattern category each cluster undergoes.

**4.1.1 The down-regulated pattern –sarcopenia signature: direct and positive connection between sarcopenia signs and 24 genes that exhibits a down-regulated pattern. Genes are involved in developmental and locomotion.**

The down-regulated pattern can be depicted in the cluster G6 and the 2 clusters that merge/split from G6: clusters G4 and G3. The main cluster G6 with down-regulated pattern has 24 genes, which is more than 50% of the genes in the list of 42 genes. A down regulation of expression of these 24 genes with age might be a direct reflection of the "age-related" loss of muscle mass leading to muscle weakness. In this sense, the 24 genes might be the representative genes for sarcopenia signature. The genes can be found in Table 1 highlighted in blue.

The common biological theme of the 24 genes is involvement in determination of adult life, locomotion, positive regulation of growth rate, larval developmental. As mentioned the down-regulated pattern of cluster G6 is maintained in clusters G3 and G4 as well as the common biological theme. Below are the down-regulated cluster patterns for the clusters G3 with the highest stability of size 13 and stability 5. The intermediate cluster G4 has size 15 and stability 3. The same common biological theme as in cluster G6 is preserved in G3 cluster also

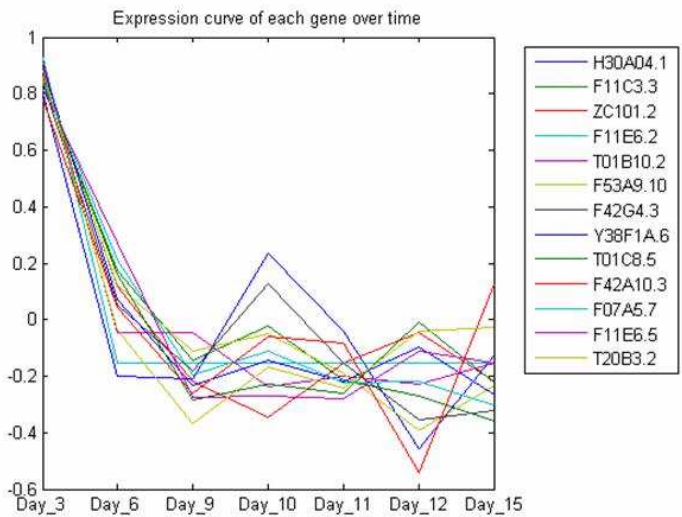

Legend: x –axis: time points, y-axis: gene expressions., log2 was applied.

Fig. 3 **G3 cluster pattern: stability 5, size 13**

Also, below is cluster G4 pattern.

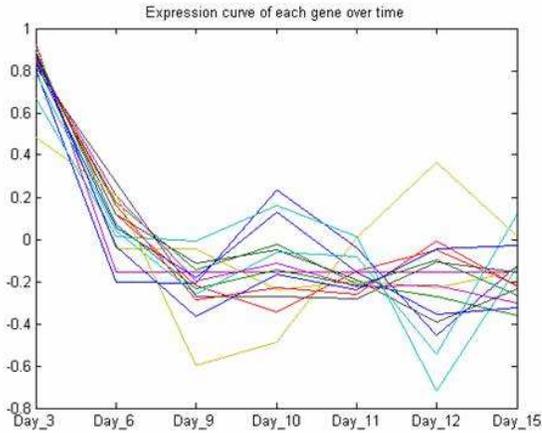

**Legend:** x –axis: time points, y-axis: gene expressions., log2 was applied.

**Fig. 4 G4 cluster pattern: stability 3, size 15.**

The emergent biological theme for the down-regulated pattern genes is the involvement in developmental and locomotion. Given our finding of the 24 genes, with a decrease in gene expression, that have a role in locomotion and that the sarcopenia phenotype is defined by signs of decreasing in locomotion functions might be that this 24 genes could play an important role in sarcopenia. The fact that the phenotypic outcome over time of this nematode shows a clearly slow movement with age and that we notice a decrease in gene expression with age of the 24 genes involved in locomotion might be an expression of a direct and positive connection between sarcopenia signs and the 24 genes.

**4.1.2 Oscillatory down-regulated day 10 pattern:** is found in 3 clusters G8,G9,G2 of very small size of 4 and 2 genes but of very high stability: 35, 12 and 5. The cluster G2 below is the cluster with highest stability among the 3.

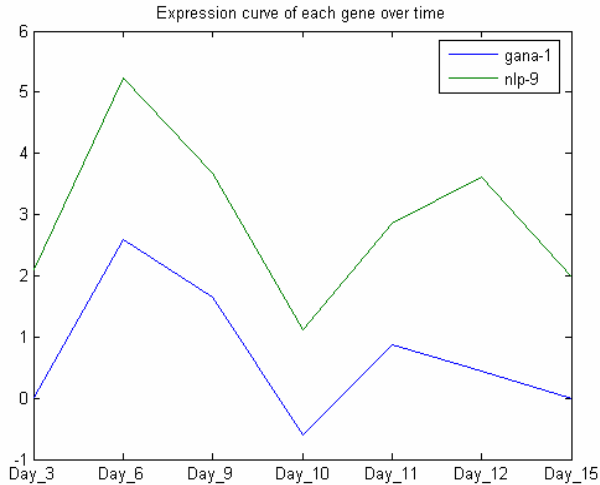

**Fig. 5 G2 cluster stability 35: x –axis: time points, y-axis: gene expressions., log2 was applied.**

The genes in this 3 clusters have in common the fact that are expressed in body wall muscle; 2 of genes: *mtm-5* and *hen-1* are expressed in pharyngeal muscle. Involved in metabolic process is *gana-1* which is also orthologous of human galactosidase. *mtm-5* has no obvious function in RNAi assays and *nlp-9* has no description of its function so far.

All 4 genes in cluster G9 exhibit an oscillatory gene expression pattern with a down pattern in gene expression at midlife time point, day 10.

**4.1.3 Up-regulated pattern:** can be noticed in the main cluster G7 and then in the cluster G5 which splits from G7. Cluster G7 has size 14 and stability 20. The cluster G5 has size 11 and stability 3.

Bellow the G7 cluster pattern can be seen.

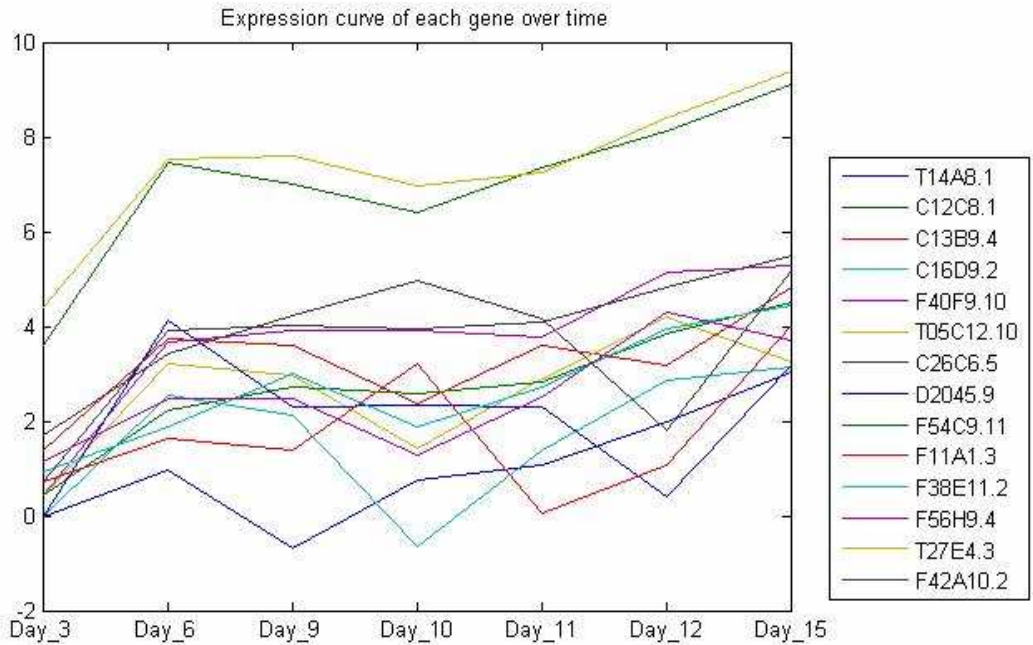

Fig. 6 G7 cluster of size 14 up-regulated pattern: x –axis: time points, y-axis: gene expressions., log2 was applied.

Most of the genes in cluster G7 are responsible for mutants defective in locomotion. Besides G7 contains *daf-12,* which encodes a member of the steroid hormone receptor superfamily homologous to the vitamin D receptor that affects dauer formation downstream of the TGF- and insulin signaling pathways. More important for the sarcopenia signature *DAF-12* together with *DAF-16* affects gonad-dependent adult longevity.

In G7 another human homolog can be found: *nfm-1,* which encodes a merlin/schwannomin (NF2), that when mutated leads to neurofibromatosis.

**The genes in cluster G7 might be required to have an increase in gene expression over time as to diminish or delay the sarcopenia signs.** A decrease in gene expression over life span of this nematode for the genes in the cluster G7 might induce more accentuated sarcopenia signs. On the other hand, an increase in gene expression over life span of the *C. elegans* for the down-regulated genes in

cluster G6 the main cluster with a down regulated pattern  might induce early signs of sarcopenia.

So far we discussed the study done on a list of genes identified (using AQL language) to be expressed in muscle cells.

**We identified 24 genes with a down –regulated pattern that might have a direct connection with the sarcopenia process. These genes are mainly in the cluster G6.  Also we identified 14 genes with an up-regulated pattern, found mainly in cluster G7. These genes might be required to have an increase in gene expression over time as to diminish or delay the sarcopenia signs or might have actually an opposite effect therefore the delay in sarcopenia signs appearances to happen if their expression would be decreasing. If this would be the case then this 14 genes are like 'leftover genes' for the sarcopenia process.**

 In the next section we performed a detailed study on a list of genes muscle-related which was compiled using Gene Ontology database. We are discussing this study in the next section.

## 4.2 Gene lists specifically involved in biological process, molecular function, or cellular component identified as muscle related.

4.2.1 Compiling gene lists involved in specific biological, cellular or functional muscle related processes.

To compile such a muscle related gene list I used Gene ontology database. The Gene Ontology (GO) project is a collaborative effort to address a consistent description of gene products in different databases.

The GO project has developed three structured controlled vocabularies (ontologies) that describe gene products in terms of their associated biological processes, cellular components and molecular functions

In this sense, using Gene Ontology database (GO-term) I searched for genes involved in biological process, molecular function, or cellular component muscle- related.

I identified genes as major structural muscle genes: myosin or actin or genes involved in biological processes as muscle contractions see Table (2) below. We keep in mind that for WormBase, GO annotation is currently a "work in progress" therefore, we've used this list just as a "guiding" list and show no surprise when some searches results were incomplete. Particularly, for cellular component, where a protein is localized within a cell, the search of genes expressed in muscle cells using AQL language outputs many more genes. The advantages of using GO-term search is that it outputs genes based on the involvement in the biological process or molecular function as known in literature.

   The number of genes found using GO_term search was 151 genes, but since the same genes were part of several biological or molecular processes, muscle involved, the total number of unique such genes was edited to 117 genes. Out of this, on our array we found a total of 95 genes, which are part of various biological processes as muscle development-the process whose specific outcome is the progression of the muscle over time, from its formation to the mature structure; or even more specific pharyngeal muscle development- the process whose specific outcome is the progression of the pharyngeal muscle over time, from its formation to the mature structure. See Table (2) for such genes.

**Table 2: genes identified using GO-term: gene muscle-related involved in development and contraction.**

| Biological process: **muscle development:** | | |
|---|---|---|
| act-1 | T04C12.6 | An actin that affects body wall and pharyngeal muscle |
| eat-1 | T11B7.4 | eat-1 encodes a homolog of mammalian |
| hlh-1 | B0304.1 | hlh-1 encodes a basic helix-loop-helix (bHLH) |
| mlc-1 | C36E6.3 | encodes a muscle regulatory myosin light chain |

| syd-2 | F59F5.6 | syd-2 encodes alpha-liprin, |
|---|---|---|
| tmd-2 | C08D8.2 | |
| unc-52 | ZC101.2 | The unc-52 gene encodes perlecan, a protein orthologues |
| unc-89 | C09D1.1 | |

Biological process: **pharyngeal muscle development**

| eff-1 | C26D10.5 | The eff-1 gene encodes a novel, type I transmembrane protein |
|---|---|---|
| glp-1 | F02A9.6 | glp-1 encodes an N-glycosylated transmembrane protein |
| pop-1 | W10C8.2 | pop-1 encodes an HMG box-containing protein |

**myosin:**

biological process: **muscle contraction:**

| egl-2 | F16B3.1 | egl-2 encodes a voltage-gated potassium channel |
|---|---|---|
| itr-1 | F33D4.2 | itr-1 encodes a putative inositol |
| jph-1 | T22C1.7 | jph-1 encodes a junctophilin, |
| pat-10 | F54C1.7 | pat-10 encodes body wall muscle troponin C, |
| twk-18 | C24A3.6 | twk-18 encodes one of 44 C. elegans TWK |
| unc-26 | JC8.10 | |
| myo-2 | T18D3.4 | myo-2 encodes a muscle-type specific myosin heavy |
| myo-3 | K12F2.1 | myo-3 encodes MHC A, the minor isoform of MHC |

Other muscle related genes were found to be implicated in biological process as muscle cell fate specification, which is the process by which a cell becomes capable of differentiating autonomously into a muscle cell in an environment that is neutral with respect to the developmental pathway. Interestingly is that upon specification, the cell fate can be reversed as for example: mls-1 which encodes a T-box transcription factor orthologous to members of the Tbx1 subfamily of T-box transcription factors. MLS-1 is required for fate specification of the eight nonstriated uterine muscle cells generated during postembryonic development. Also ectopic expression of MLS-1 is sufficient for uterine muscle specification in other mesodermal lineages. Besides mls-1 reporter gene expression is detected in uterine progenitors and differentiated uterine muscles, type 2 vulval muscles, the left and right intestinal muscles, and the anal depressor muscle.

Another important biological process I included in my search was muscle contraction which is a process leading to shortening and/or development of tension in muscle tissue. Muscle contraction occurs by a sliding filament mechanism whereby actin filaments slide inward among the myosin filaments. Major structural muscle genes as myosin and actin.

Besides muscle related genes involved in biological processes, using GO term I identified also genes involved in various biological functions as actin cytoskeleton organization and biogenesis by which we understand the assembly and arrangement of cytoskeletal structures comprising actin filaments and their associated proteins. See table 3 bellow for such genes.

| act-1 | T04C12.6 | An actin that affects body wall and pharyngeal mus |
| act-4 | M03F4.2 | An actin that is expressed in body wall and vulval |
| cap-1 | D2024.6 | cap-1 encodes an F-actin capping protein alpha sub |
| cap-2 | M106.5 | The beta subunit of actin capping protein that reg |
| cyk-1 | F11H8.4 | The cyk-1 gene encodes a homolog of Drosophila |
| fhod-1 | C46H11.11 | |
| fhod-2 | F56E10.2 | |
| fozi-1 | K01B6.1 | K01B6.1 encodes a protein with a zinc-finger domain |
| pfn-1 | Y18D10A.20 | |
| pfn-3 | K03E6.6 | |
| tag-268 | F58B6.2 | |
| unc-53 | F45E10.1 | UNC-53 encodes at least five large (~1200-1600 residues |
| | F15B9.4 | |
| | F56E10.3 | |
| | Y48G9A.4 | |

**Table 3 genes involved in actin cytoskeleton organization and biogenesis**

Another muscle related biological function would be actin filament organization by which we understand control of the spatial distribution of actin filaments. This includes organizing filaments into meshworks, bundles, or other structures, as by cross-linking. Genes involved in this process would be: *ced-12, die-1, wve-1*

All 117 genes found based on the search using GO-term involved in various biological, molecular or cellular function are presented  in several tables in Appendix A -at the beginning of each table I give the definition of the process in which the respective genes are involved. Note that some genes have no description in these tables as GO-term data base is still a work in progress in *C. elegans* community.

### 4.2.2 Clustering the gene lists compiled in previous section

In order to identify the gene expression pattern of the 95 muscle related genes found on our chips, we normalized and clustered the genes using the SPC algorithm. For a description on the methods see Chapter 1.  Appendix B has the list of 95 genes as well as gene members of clusters G8 and G10.

We obtained 10 clusters annotated from G1-G9 where G1 contains the entire data to be clustered. The clusters are classified based on size and stability as can be seen in Table 5 bellow:

**G1** Size=95

| G1(S1) | G2 | G3 | G4 | G5 | G6 | G7 | G8 | G9 | G10 |

| **G2**  **Stability=9 Size=5** |
|---|
| **G3**  **Stability=3 Size=4** |
| **G4**  **Stability=3 Size=10** |
| **G5**  **Stability=4 Size=30** |
| **G6** **Stability=5 Size=5** |
| **G7** **Stability=3 Size=7** |
| **G8** **Stability=13 Size=42** |
| **G9** **Stability=3 Size=19** |
| **G10** **Stability=8 Size=47** |

As in previous analysis we compiled the results of the clustering analysis for an easy access in a web based design. The entire informational content of the web based clustering design is displayed graphically or in tables.

The hierarchical organization found in the data based on SPC algorithm is presented in the dendogram bellow.

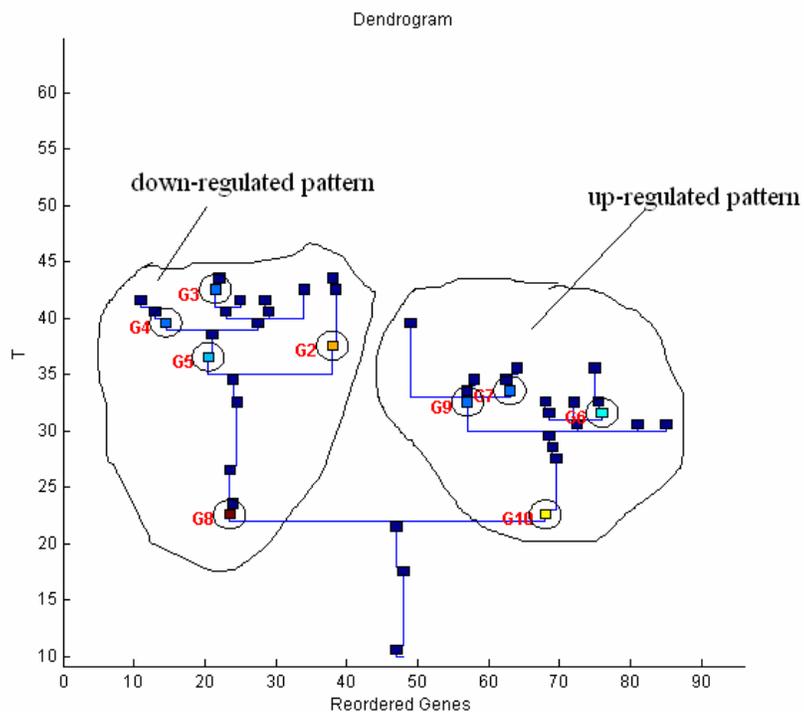

Fig. 7 Dendogram with gene nodes/clusters and patterns

We depicted 2 main patterns in the clustered data. The dendogram above shows the 10 clusters and the pattern category in which the clusters enter.

**Most of the structural genes are in the down-regulated pattern** found in the cluster G8 of size 42 and stability 13 and all the clusters which split from it: G5,G4,G3,G2.

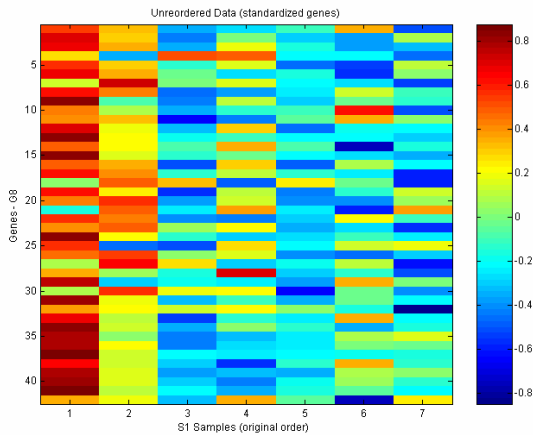

**x-axis**-7 time points: 1/day3; 2/day6; 3/day9; 4/day10; 5/day11; 6/day12; 7/day15;**y-axis**-42 gene expressions; color bar: from red-high gene expression too dark blue-low gene expressions.

**Fig. 8 Heat map cluster G8 -down-regulated pattern**

As can be seen in the heat map above of the cluster G8, the gene expression over the first 2 time points (day3, day 6) are higher expressed than the gene expression for the rest of the time points: day 9,10,11,12,15. At the same time it should be mentioned that the gene expression for the rest of the time points 9,10,11,12,15 have an relative steady down -regulated pattern. This pattern is maintained in all clusters that merge from cluster G8 and mentioned before: G5,G4,G3,G2.

Below can be seen the gene expression pattern for some of the clusters that merge from cluster G8: clusters G4 and G3

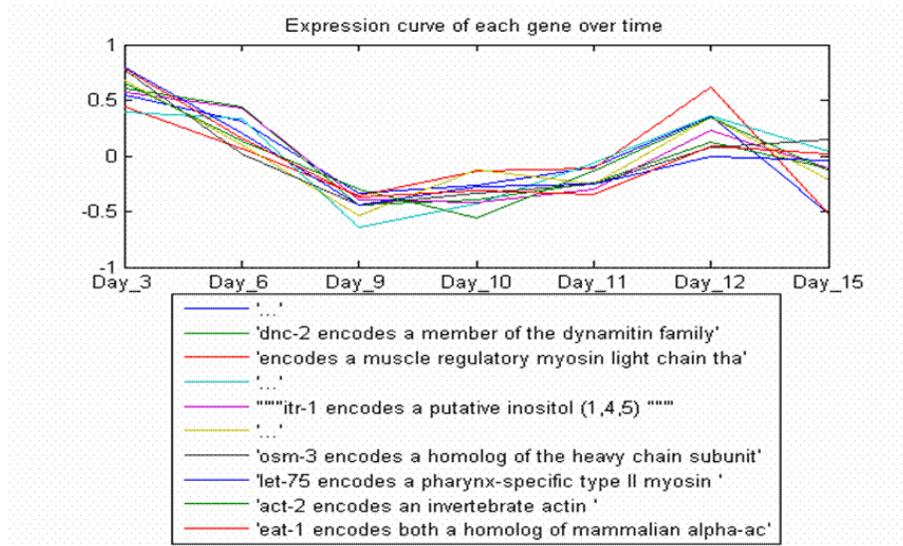

x –axis: time points, y-axis: gene expressions., log2 was applied.

Fig. 9 G4 cluster gene expression pattern over time points.

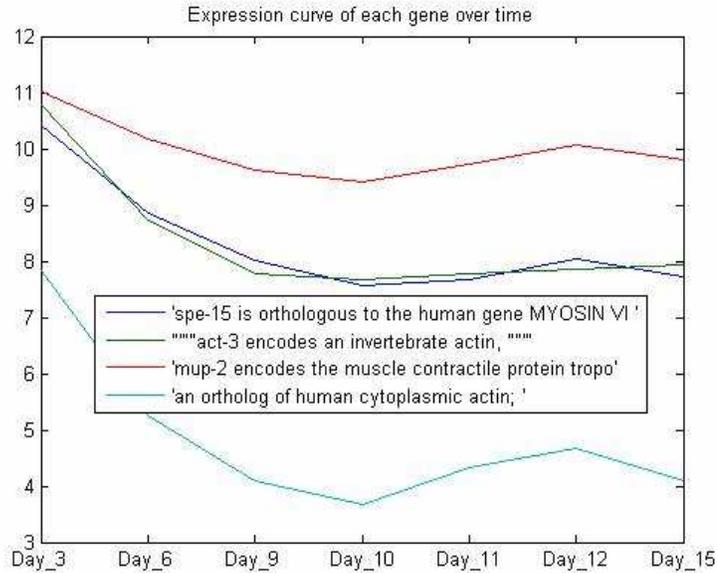

x –axis: time points, y-axis: gene expressions., log2 was applied.

Fig.10 **G3 cluster: gene expression pattern over time points**

As mentioned, most of the structural genes are in down-regulated clusters pattern. S**tructural muscle related genes show 'positive connection' with the sarcopenia phenotype, meaning they have the down-regulated gene expression pattern one might expect for a sarcopenia signature.**

In the **up-regulated pattern** found in G10 **most of the unc genes** and some human homologues genes **are included**. The emerging clusters from G10 are: G9,G7,G6. They do maintain same pattern as G10 cluster.

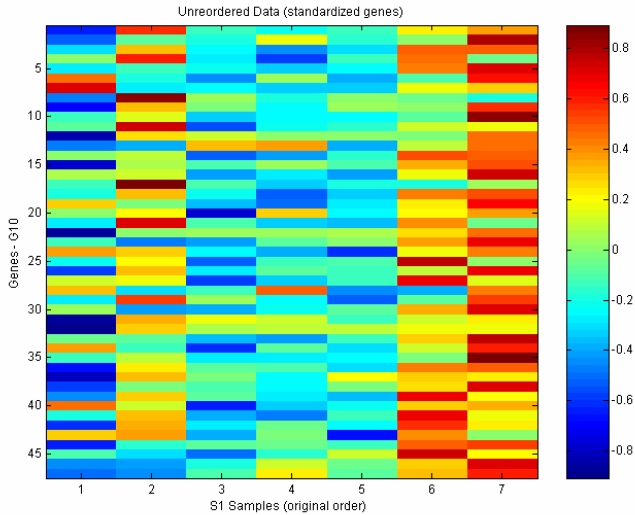

**Legend:** **x-axis**-7 time points: 1/day3; 2/day6; 3/day9; 4/day10; 5/day11; 6/day12; 7/day15; **y-axis**-42 gene expressions; color bar: from red-high gene expression too dark blue-low gene expressions.

**Fig. 11 Heat map-cluster G10 down-regulated pattern size 47, stability 13.**

In the heat map of the cluster G10, gene expression for the last 2 time points (day12, day 15) are expressed higher than the gene expression for the rest of the time points: day 3,6,9,10,11.

It should be mentioned that the gene expression for the rest of the time points day 3,6,9,10,11 show an almost steady down-regulated pattern. This pattern is maintained in all clusters that merge from cluster G10: G9,G7,G6.

Below the gene expression pattern for the clusters: G7 and G6 can be seen.

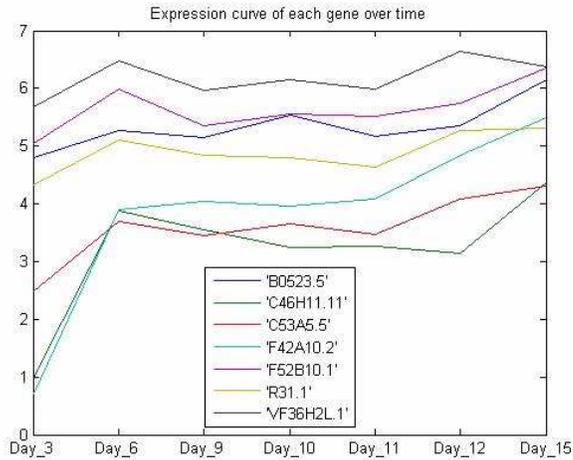

x –axis: time points, y-axis: gene expressions, log2 was applied.

**Fig. 12 G7 up-regulated cluster: gene expression pattern over time points**

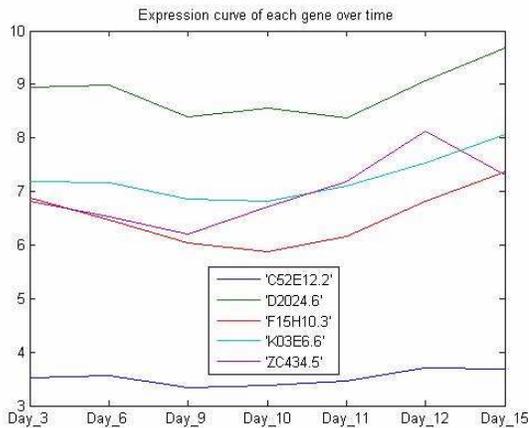

x –axis: time points, y-axis: gene expressions, log2 was applied.

**Fig.13 G6 up-regulated cluster: gene expression pattern over time points.**

The fact that in up-regulated pattern clusters I identified mostly unc genes might be again a signature of sarcopenia. Might be that the unc genes for the fitness of the muscle acts as 'leftover' genes. In this sense, mutants with a decrease in gene expression for the genes in the up-regulated cluster G10 might slow the sarcopenia signs.

**To conclude the GO-term list analysis, the gene lists specifically involved in biological process, molecular function, or cellular components considered to be muscle related can be classified in two categories of gene expression pattern : of up-regulated and down-regulated genes. In the down-regulated group we found more structural muscle related genes like actin or myosin by difference with the up-regulated group where we can see more of unc-related genes as well as human homologues genes.**

**The decrease in gene expression for structural genes might help promoting sarcopenia signs. In the same time the increase in gene expression for unc genes identified in cluster G10 as far as concerning sarcopenia, might be a sign of 'leftover' genes, in the sense that mutants with a decrease in gene expression for genes in cluster G10 might improve the the reduction in muscle mass and any other sarcopenia signs in general.**

### 4.3 Analysis of genes expressed in young muscle

In 2002 an experiment performed by Kim's group identified gene expressed in C. elegans muscle (see Kim, et al. Nature 2002);

In this experiment a poly A binding protein was expressed only in muscle. Muscle messages were then isolated by imunoprecipitation. They used DNA microarrays to analyze the ratio of the mRNA enriched by co-immunoprecipitation with FLAG::PAB-1 relative to the mRNA present in the starting cell-free extract. Fluorescently-labeled probes were then hybridized to DNA microarrays containing 90% of the 19,733 genes currently estimated in the *C. elegans* genome. L1 larvae were studied with 6 repeats for ~19000 genes.

As statistical method used, they've computed a ranking for all genes and a percentile rank for every gene from the 6 repeats. Then, the percentile rank of enrichment for every gene from the six repeats was averaged together.

They considered that genes that are not enriched by mRNA-tagging should have an average percentile rank of about 50%, while genes expressed in muscle should have a rank significantly higher.

After that they performed a Student's t-test and identified 1364 genes that are significantly enriched in the muscle mRNA-tagging experiments for p<0.001. See the graphs bellow.

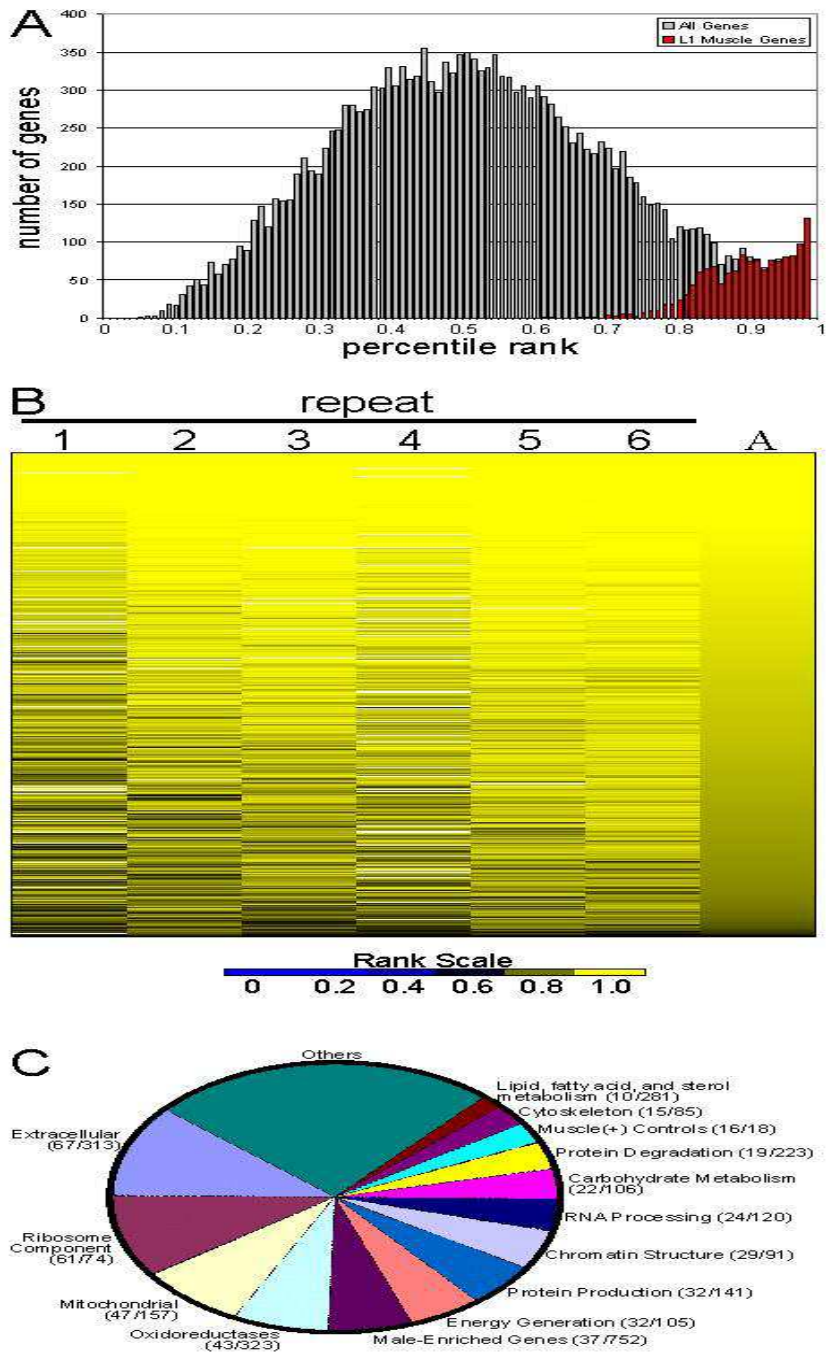

**Fig. 14- from Kim et. al. Nature 418: 975-979: 2002;**

In the pie chart above can be seen the biological classification of the 1364 genes identified by t-test and enriched by co-immunoprecipitation with FLAG::PAB-in the L1 stage of life. **The 1364 genes are part of almost every known biological function**.

4.3.1 Analysis of muscle enriched genes  identified by Kim et.al;  changes in adult life

 Out of 1364 genes from Kim et. al.  experiment we identified 1187 genes on our chips.

We looked for gene expression patterns in the list of 1187 genes and we analyzed the  intersection between 1187 muscle expressed genes and the list of 2000 genes from our experiment which have the highest variation across tie points. We identified 111 genes at the intersection of the 2 lists.  These 111 are genes expressed in L1 muscle that show greatest variance sometime during adulthood in our experiment.

   When we analyzed the gene expression patterns in the list of 1187 using SPC algorithm we identified 28 clusters. (For details on the clustering algorithm see Chapter 1 as well as  M. Blatt, S. Wiseman and E, Domany, Neural Computation (1997)).

    The Fig. 4 below shows the dendogram that reflects the clustering hierarchy identified in the data as well as the heat map graph for the 1187 genes as found on our arrays.

Blue means low gene expression, red is high gene expression.

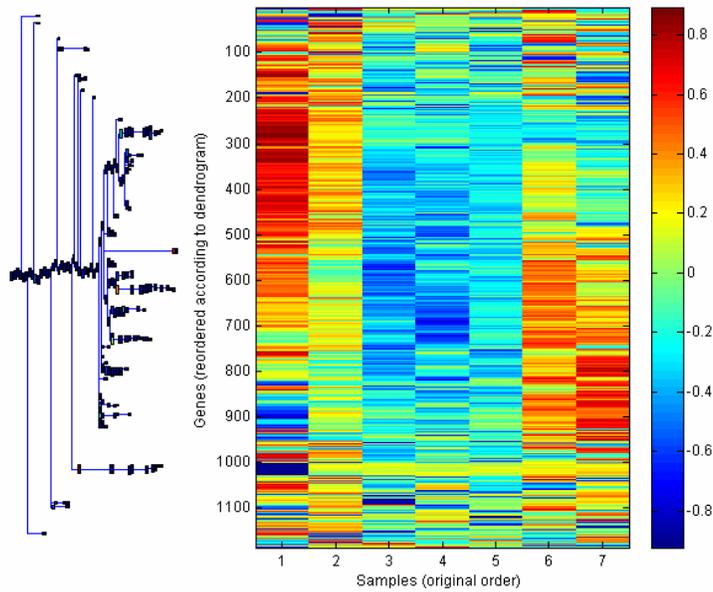

Fig. 15 Heat map and dendogram of the 1187 genes corresponding to the 1364 genes identified by Kim et. al. as expressed in muscle

Given that the 1364 genes identified by Kim group are part of almost every known biological function came as no surprise that same biologically consistency we identified in our list of 1187. In the same time similar patterns identified when analyzed the 2000 list of genes which vary most in our data were found when analyzed the list of 1187 genes.

Fig. 16 bellow shows the dendogram with hierarchical organization of the clusters and the 5 category patterns they enter.

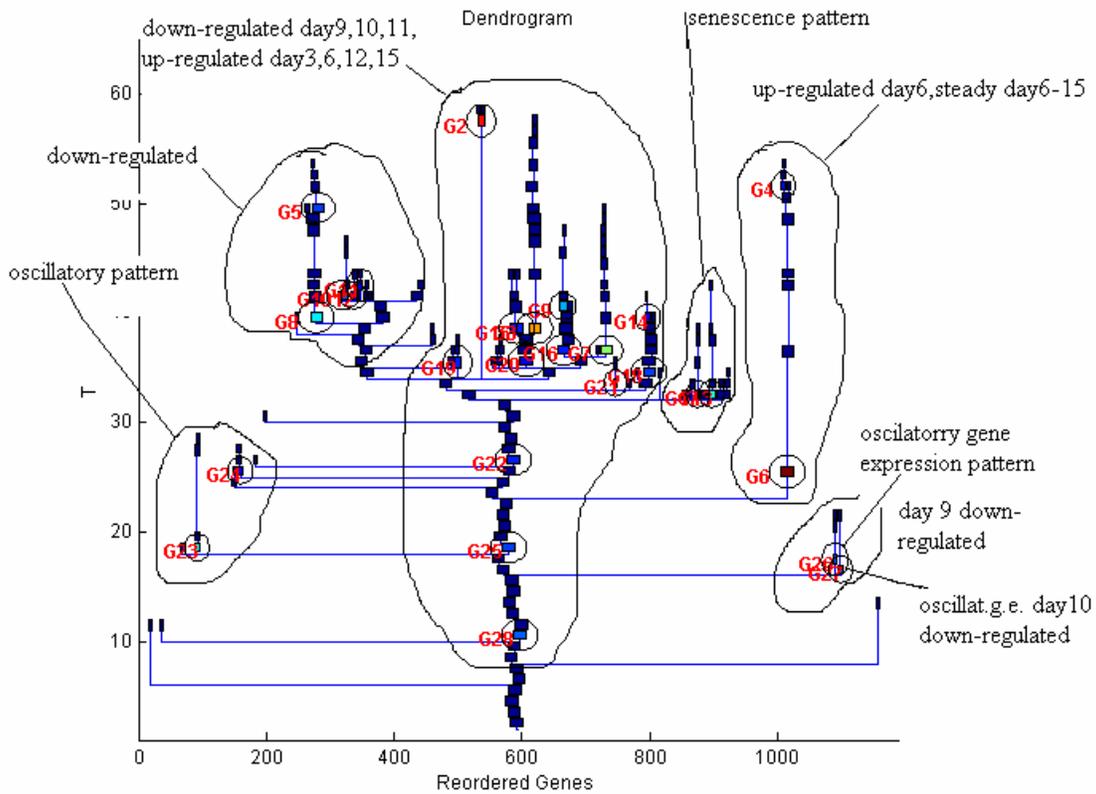

Fig. 16 dendogram – 28 clusters and the 5 category patterns they enter

The 28 clusters can be identified in five gene expression patterns:

1) oscillatory pattern,

2) down-regulated pattern,

3) down-regulated mid life time points: day9,10,11 and up-regulated gene expression pattern for day3,6,12,15

4) "Senescence pattern"-oscillatory low expressed day3-12, up-regulated day12-15

5) a)Up-regulated pattern

   b)"Developmental pattern"-steady state day6-15, up-regulated day3-6

**1) oscillatory gene expression pattern ( 4 clusters with this pattern)- mostly genes with unknown protein function and growth defect.**

Cluster G26 has size 8 and stability 6 with a down-regulated peck at day 9. See graph below.

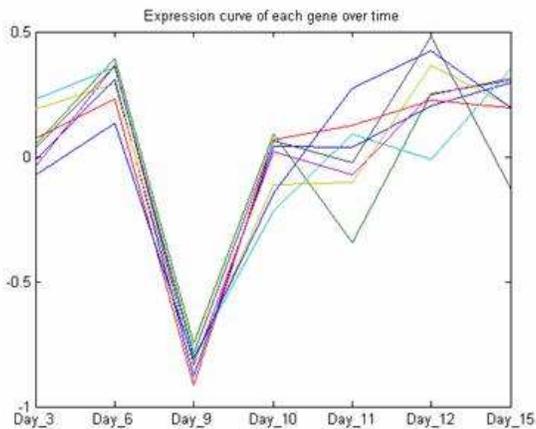

**Fig. 17- G26 cluster**

In this cluster we have many unknown function genes, genes involved in growth defect as RNAi phenotype and cuticulin component related gene. See below Table with cluster members.

| 1 | 'F41G4.7' | F41G4.7 Protein of unknown function |
|---|---|---|
| 2 | 'C27B7.6' | C27B7.6 Member of the protein phosphatase protein family |
| 3 | 'T06A10.1' | T06A10.1/mel-46 ; RNAiphenotype:growth defect |
| 4 | 'W01H2.2' | W01H2.2 Protein of unknown function |
| 5 | 'F55C12.4' | F55C12.4 Protein of unknown function |
| 6 | 'F53F1.1' | F53F1.1 Protein with strong similarity to C. elegans cut-1 (Cuticulin component) |
| 7 | 'Y71H9A.3' | Y71H9A.3 Member of the stomatin protein family |
| 8 | 'H13N06.2' | H13N06.2 Protein with weak similarity to C. elegans mup-4 (Member of the EGF-repeat protein family) |

**Table 4: G26 cluster members.**

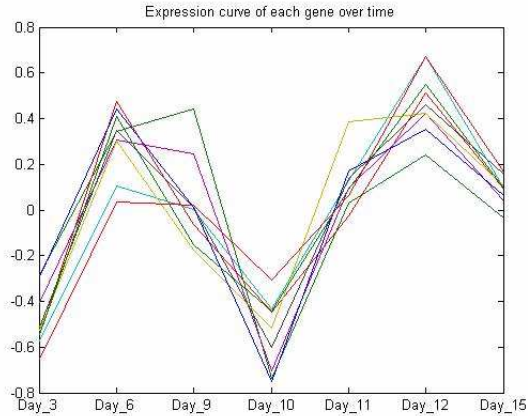

Fig. 18 **G27 cluster, size 10, stability 6.**

Again in this cluster we have mostly genes that have an unknown protein functions, a slow growth gene, and a transcription factor gene. See Table x bellow for cluster members in the G27 cluster.

| 1 | 'M02F4.1' | "M02F4.4 Protein of unknown function, has strong similarity to C. elegans R04D3.3 " |
|---|---|---|
| 2 | '<u>ZK54.1</u>' | is on Chromosome X;Protein of unknown function; has SNP's ; |
| 3 | 'M02F4.1' | "M02F4.4 Protein of unknown function, has strong similarity to C. elegans R04D3.3 " |
| 4 | '<u>W08A12.2</u>' | W08A12.2 Protein of unknown function |
| 5 | '<u>F28A12.3</u>' | "F28A12.3 Protein of unknown function, has strong similarity to C. elegans F35C5.11 " |
| 6 | 'ZC334.2' | "ZC334.2 /<u>ins-30</u>;Protein of unknown function, has weak similarity to C. elegans ZC334.3 " |
| 7 | 'C08E3.1' | C08E3.1 Member of an uncharacterized protein family |
| 8 | 'E03A3.3' | E03A3.3/<u>his-69</u>; Member of the histone H3 protein family; RNAi phenotype-slow growth |
| 9 | 'C33D12.1' | "C33D12.1/<u>ceh-31</u>    Homeodomain    transcription    factor,    has similarity over 121 amino acids to D. melanogaster B-H1 (BarH1) homeodomain transcription factor " |
| 10 | '<u>C25G4.7</u>' | is on Chromosome IV; "C25G4.7 Protein of unknown function, has strong similarity to C. elegans ZK973_14.D "; has SNP's ; |

**Table 5- G27 cluster members**

Same oscillatory pattern is maintained in the clusters G23 and G24. **In this clusters can be found also member of the chaperonin complex protein family, heat shock proteins.**

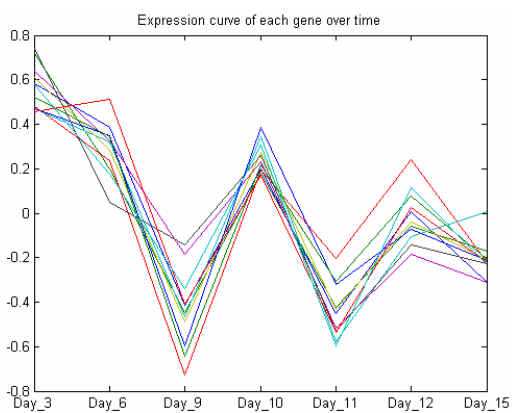

Fig. 19 Cluster G23-0x axis- time points; 0y axis- normalized gene expression

**2) down-regulated pattern: pattern seen in 5 clusters-mostly collagen genes**

**G8**

Down-regulated

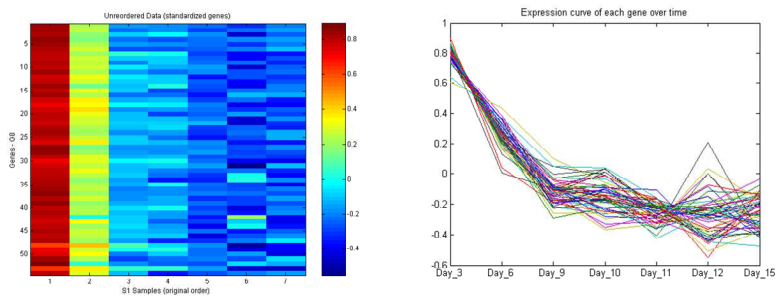

**Fig.20 Cluster G8-size 54, stability 10, right: heat map graph, left: normalized gene expressions.**

Same pattern is maintained also in clusters G5,10,11,12.

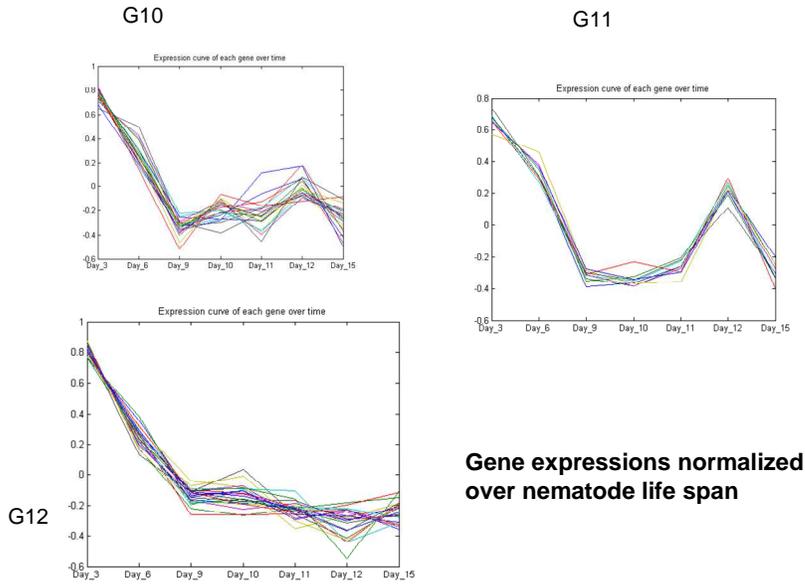

**Fig. 21 Cluster pattern for G12,G10,G11**

The cluster G8 the biggest size cluster with a down-regulated pattern has mostly collagen related genes, a few genes with uncharacterized protein function and several structural muscle related, no more than 4 genes out of 54 gene expressions, the size of the G8 cluster.

3) **down-regulated mid life time points: day9,10,11 and up-regulated gene expression pattern for day3,6,12,15-pattern found in 15 clusters-mostly ribosomal related genes.**

This pattern can be seen in 15 clusters out of 28 clusters we identified. ~ 50% of clusters have a low expression pattern for day 9,10,11 the mid life time span of this organism. The overwhelming theme in these clusters is the ribosomal proteins and genes with unknown protein function. The heat map of the cluster G20 which is representative for this pattern is presented bellow:

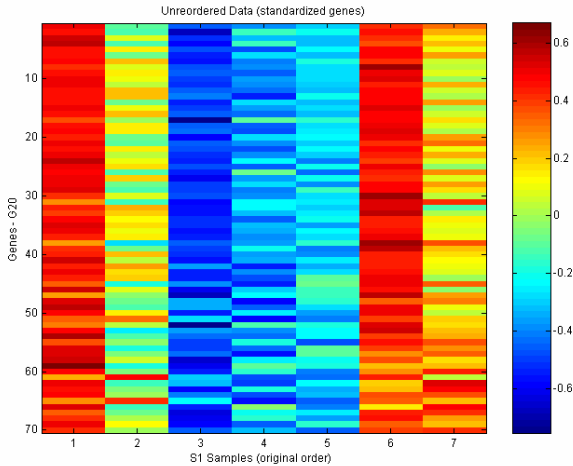

Fig. 22 **Heap map of the G20 cluster; x-axis**-7 time points: 1/day3; 2/day6; 3/day9; 4/day10; 5/day11; 6/day12; 7/day15; **y-axis**-42 gene expressions; color bar: from red-high gene expression too dark blue-low gene expressions.

The lower gene expression for the time points 3,4,5 = day9,10,11 can be clearly seen in the heat map above. Same pattern is maintained in the rest of 14 clusters.

**4) Senescence pattern- oscillatory low expressed- day3-12, up-regulated day12-15. Pattern noticed in one cluster-genes mostly involved in growth defect.**

Cluster G17 has this pattern**.**

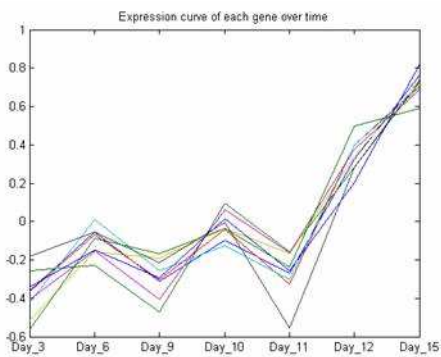

Fig. 23 **Cluster G17 size 9 stability**

Cluster G17 has mostly genes involved in growth rate. See cluster members, in Table x below.

| 1 | 'F21D5.6' | F21D5.6 Protein of unknown function |
|---|---|---|
| 2 | 'C16A3.3' | "C16A3.3 Protein with strong similarity to S. cerevisiae Rrp5p, an essential protein required for processing of pre-rRNA to 18S and 5.8S rRNA "; RNAi phenotype: Egl Emb Gon Gro Lva; biological process:embryonic development gonad development growth (IMP) larval development oviposition (IMP) positive regulation of growth rate (IMP) RNA processing |
| 3 | 'F33D4.5' | F33D4.5 Protein of unknown function/ RNAi phenotype-embryonic & post-embryonic defect, lethal |
| 4 | 'C34B2.5' | C34B2.5 Protein with moderate similarity to human TTC1 (tetratricopeptide repeat domain 1);Larval Arrest-Late (L3/L4) |
| 5 | 'K08F4.1' | "K08F4.1 Protein with strong similarity to S. cerevisiae Ctf18p, which is required for chromosome transmission and maintenance of normal telomere... "; homologies with DNA helicases in H.Sapiens; function:DNA replication |
| 6 | 'ZK632.3' | ZK632.3 Member of the RIO1/ZK632.3/MJ0444 protein family |
| 7 | 'M04B2.3' | "M04B2.3 Protein with strong similarity human GAS41/Hs.4029, which is amplified in glioma cells " |
| 8 | 'F16A11.2' | F16A11.2 Member of the uncharacterized UPF0027 protein family; RNAi phenotype: slow growth; developmental delay |
| 9 | 'ZK930.2' | ZK930.2 Protein of unknown function |

**Tabele 6. G17 cluster members.**

**5) a) Up-regulated pattern**

**- pattern found in one cluster-size 14-same biological content as G17-mostly genes involved in growth defect**

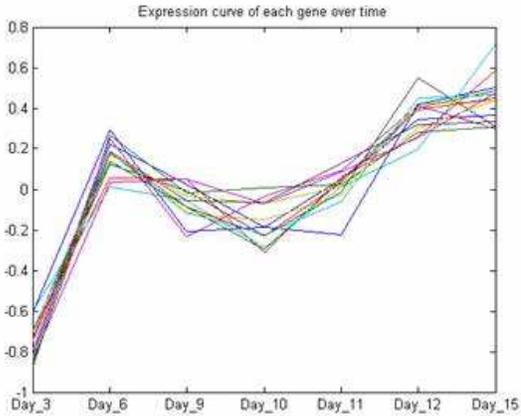

Fig. 24 **Cluster G13 size 14, stability 11.**

Biological theme is the same as for cluster G17. Most of the genes are involved in growth defect, **developmental delay, embryonic &postembryonic defect, larval arrest in L3/L4 stage** as can be seen in Table x bellow. Cluster G13 has size 14 and stability 11.

| 1 | 80 | 'F54F2.7' | F54F2.7 Protein of unknown function; RNAi phenotype: ==growth defect== |
|---|---|---|---|
| 2 | 453 | 'E04F6.9' | "E04F6.9 Protein of unknown function, has moderate similarity to C. elegans E04F6.8 " |
| 3 | 540 | 'W06E11.2' | W06E11.2 Protein of unknown function;RNAi phenotype: ==growth defect== |
| 4 | 553 | 'F19B2.5' | F19B2.5 Protein with strong similarity to C. elegans F19B2.G |
| 5 | 599 | 'B0212.1' | "B0212.1 Protein of unknown function, has strong similarity to C. elegans F14B8.5 " |
| 6 | 613 | 'F42H10.4' | F42H10.4/cal-2; Member of the LIM domain containing protein family; |
| 7 | 672 | 'C18E9.1' | C18E9.1 Member of an uncharacterized protein family; RNAi phenotype: ==embryonic defect== |
| 8 | 718 | 'R05D11.3' | "R05D11.3 Putative nuclear transport factor, has similarity to human NTF-2 (nuclear transport factor 2) ==embryonic defect== |
| 9 | 730 | 'F09F7.7' | F09F7.7 Protein of unknown function |
| 10 | 961 | 'T27E4.1' | T27E4.1 Protein of unknown function; is part of cluster aging of 164 genes, see Lund et.al '02 |
| 11 | 1066 | 'C54E10.6' | "C54E10.6 Small protein containing a CHROMO (CHRromatin Organization MOdifier) domain and a coiled-coil region, has similarity to C. elegans CEC-... " |

| 12 | 1083 | 'R05G6.10' | R05G6.10 Protein containing an N-terminal RasGEFN domain and a C-terminal RasGEF domain; has similarity over C-terminal half to CDC25-like GDP/G...biological process: intracellular signaling cascade |
|----|------|-----------|------|
| 13 | 1088 | 'F45E12.6' | F45E12.6 Protein of unknown function |
| 14 | 1120 | 'T22B11.4' | "T22B11.4 Protein of unknown function, has weak similarity to human kinase scaffold protein GRAVIN (Hs.788) " |

Table 7. cluster G13 gene  members

5)  b)  **Developmental up-regulated pattern"-steady state day6-15, up-regulated day3-6.Two clusters are sharing this gene expression pattern.**

Cluster G6 and G4 share this pattern.

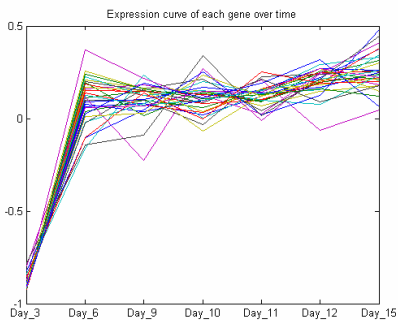

**Fig. 25 cluster G6 size 29, stability 28, normalized gene expressions**

The cluster G6 has size 29 and very high stability 28.

The gene expression pattern is characterized by high gene expression for day6-day15 and low expression at day3.

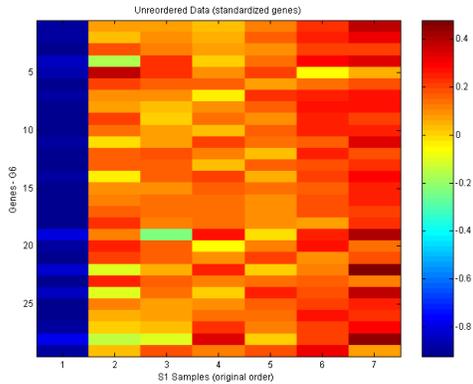

Fig. 26 **Heat map of cluster G6.**

According with expression pattern the gene cluster members might be of importance for the entire life of *C. elegans* right after development time, day3. Interestingly this cluster has a few genes related with proteins known to be human similar. Many of the genes have an unknown protein functions.

Below is the table with gene members in cluster G6.

| 1 | 'C40H1.5' | C40H1.5 Member of an uncharacterized protein family |
|---|---|---|
| 2 | 'F25H8.5' | "F25H8.5 Putative paralog of dur-1, protein with weak similarity to H. sapiens SNCB (synuclein, beta) " |
| 3 | 'C08F8.7' | C08F8.7 Ras-related GTP-binding protein of the ras superfamily |
| 4 | 'F44G4.6' | F44G4.6 Protein of unknown function |
| 5 | 'W01F3.3' | W01F3.3 Member of the EGF-repeat protein family |
| 6 | 'F59F5.6' | F59F5.6 Member of liprin (LAR-interacting protein) family of proteins |
| 7 | 'C13C12.1' | C13C12.1 Calmodulin |
| 8 | 'F42H10.3' | F42H10.3 Member of the src homology domain 3 protein family |
| 9 | 'F41B5.1' | F41B5.1 Protein of unknown function |
| 10 | 'C28H8.2' | C28H8.2 Protein of unknown function |
| 11 | 'F19F10.9' | "F19F10.9 Putative antigenic peptides, has strong similarity to H. sapiens SART1 gene product [squamous cell carcinoma antigen recognised by T ce... " |
| 12 | 'C17G10.6' | "C17G10.6 Protein of unknown function, contains a C-terminal ShKt (toxin) domain, has weak similarity over middle region to human TGN51 (trans-Go... " |

| 13 | 'F27D9.8' | "F27D9.8 Protein with strong similarity to human Hs.172278 protein, beta2-syntrophin " |
|----|-----------|----------|
| 14 | 'M05D6.4' | M05D6.4 Member of the esterase protein family |
| 15 | 'C18B10.3' | "C18B10.3 Protein with similarity to G-protein coupled receptors of an unnamed subfamily, no homolog found in human or D. melanogaster, may have ... " |
| 16 | 'ZC239.6' | ZC239.6 Member of an uncharacterized protein family |
| 17 | 'F43C1.3' | F43C1.3 Protein with weak similarity to S. cerevisiae HIT1 (Protein required for growth at high temperature) |
| 18 | 'C34C6.2' | C34C6.2 Protein of unknown function |
| 19 | 'K02F3.4' | "K02F3.4 Protein with weak similarity to H. sapiens CEBPG (CCAAT/enhancer binding protein (C/EBP), gamma) " |
| 20 | 'T13C2.3' | T13C2.3 Protein with weak similarity to C. elegans Y97E10AR.E gene product |
| 21 | 'C45H4.17' | Y5H2B.F Protein with similarity to cytochrome P450; putative ortholog of C. elegans C45H4.2 |
| 22 | 'VF39H2L.1' | "VF39H2L.1 Protein with weak similarity to human syntaxin 7 (STX7 ), has weak similarity to C. elegans F36F2.4 " |
| 23 | 'T14B4.6' | T14B4.6 Collagen of the collagen triple helix repeat (20 copies) family |
| 24 | 'C17H1.7' | C17H1.7 Member of an uncharacterized protein family |
| 25 | 'F34H10.1' | F34H10.1 Member of the ubiquitin protein family |
| 26 | 'Y40B10B.1' | Y40B10B.1 Member of an uncharacterized protein family |
| 27 | 'F56A3.1' | F56A3.1 Protein of unknown function |
| 28 | 'F09E8.2' | "F09E8.2 Protein containing EGF-like repeats, has weak similarity to human low density lipoprotein receptors and D. melanogaster TEN-1 (tenascin) " |
| 29 | 'F54F2.6' | F54F2.6 Protein of unknown function |

Table 8. gene members cluster G6.

The same pattern characterized by high gene expression for the time points 2-7, which corresponds to day 6-day 15 and a low gene expression day 3 (see fig. 26) is maintained in the cluster G4 which splits from G6.

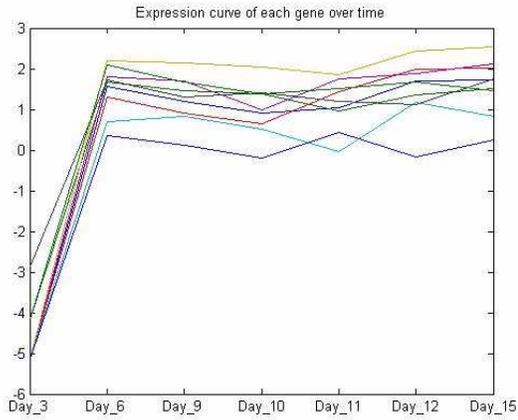

Fig. 27 **Cluster G4, size9, stability3, normalized gene expressions**

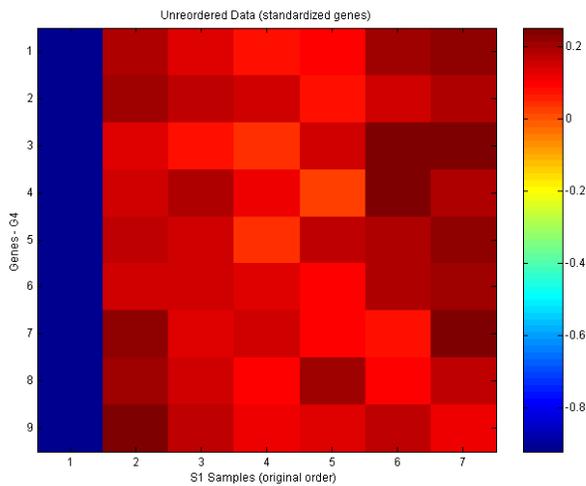

**Fig. 28 Heat map cluster G4: x-axis**-7 time points: 1/day3; 2/day6; 3/day9; 4/day10; 5/day11; 6/day12; 7/day15; **y-axis**-42 gene expressions; color bar: from red-high gene expression too dark blue-low gene expressions.

Given the expression pattern of cluster G4, the genes in this cluster might be relevant for the time right after development of *C. elegans* from day 6 up to day 15. The high gene expression pattern for most of the life of this organism might imply that this genes play an important role in the maintenance of vital functions for this nematode. This cluster has again a few genes related with proteins known to be human similar.

| 1 | 386 | 'C08F8.7' | C08F8.7 Ras-related GTP-binding protein of the ras superfamily |
|---|---|---|---|
| 2 | 525 | 'F59F5.6' | F59F5.6 Member of liprin (LAR-interacting protein) family of proteins; biological process: muscle development |
| 3 | 666 | 'C28H8.2' | C28H8.2 Protein of unknown function |
| 4 | 761 | 'C17G10.6' | "C17G10.6 Protein of unknown function, contains a C-terminal ShKt (toxin) domain, has weak similarity over middle region to human TGN51 |
| 5 | 787 | 'F27D9.8' | "F27D9.8 Protein with strong similarity to human Hs.172278 protein, beta2-syntrophin " |
| 6 | 956 | 'F43C1.3' | F43C1.3 Protein with weak similarity to S. cerevisiae HIT1 (Protein required for growth at high temperature) |
| 7 | 970 | 'C34C6.2' | C34C6.2 Protein of unknown function |
| 8 | 1005 | 'C45H4.17' | Y5H2B.F Protein with similarity to cytochrome P450; putative ortholog of C. elegans C45H4.2; biological process:electron transport |
| 9 | 1012 | 'T14B4.6' | T14B4.6 Collagen of the collagen triple helix repeat (20 copies) family |

Table 9. G4 cluster members

Given that genes that might have an important role in the life of this organism have similarities with human proteins brings again another argument for the importance of using *C.elegans* as animal model for aging studies.

Concluding the clustering analysis the majority of the genes proposed by Kim et.al. and identified on our arrays, a list of 1187 genes, can be found in 2 main pattern categories. The decreasing pattern category and the pattern category with low gene expression over mid life time of the nematode. The decreasing pattern is consistent with what we might consider a sarcopenia signature and consists of mostly collagen- related genes. This pattern is seen in 5 clusters. The pattern of low gene expression over the mid life has mostly ribosomal genes. 14 clusters out of 28 clusters, includes these, we identified from clustering the entire list of 1187 genes ~ 50% out of all clusters.  The rest of the genes enter in 3 other pattern categories of 'senescence', 'developmental' and oscillatory.

  In general, the cluster patterns found from analysis of the list of 1187 genes proposed by Kim group resemble a lot the cluster patterns we found when analyzing the clusters of  2000 gene list of the 'wild type' *C. elegans* (see Chapter 1 results).

  We might conclude that analysis of the experiment performed by Kim et.al brought us valuable information about *C. elegans* in general and development for the larval L1 of stage. For more  insights on muscle related genes and sarcopenia as a process new experiments should be designed.

### Intersection between 1187 gene list and 2000 gene list

       We've looked also at the intersection between Kim list and our list of 2000 genes (see Chapter 1) and we found 111 genes that are both expressed in L1 muscle and have high variation in gene expressions  with time.  The lists analyzed in Kim et.al. experiment are schematized in the Fig. x below.The pie chart in Fig. x shows the biological composition in the list of 111 genes. 41% have collagen related genes, 11% are genes with human homologies, 12% are muscle related genes, and, 36% genes with unknown protein related function.

**Analysis performed using   data from  Kim, et.al** *2002. Chromosomal clustering of muscle-expressed genes*  Nature **418**: 975-979;

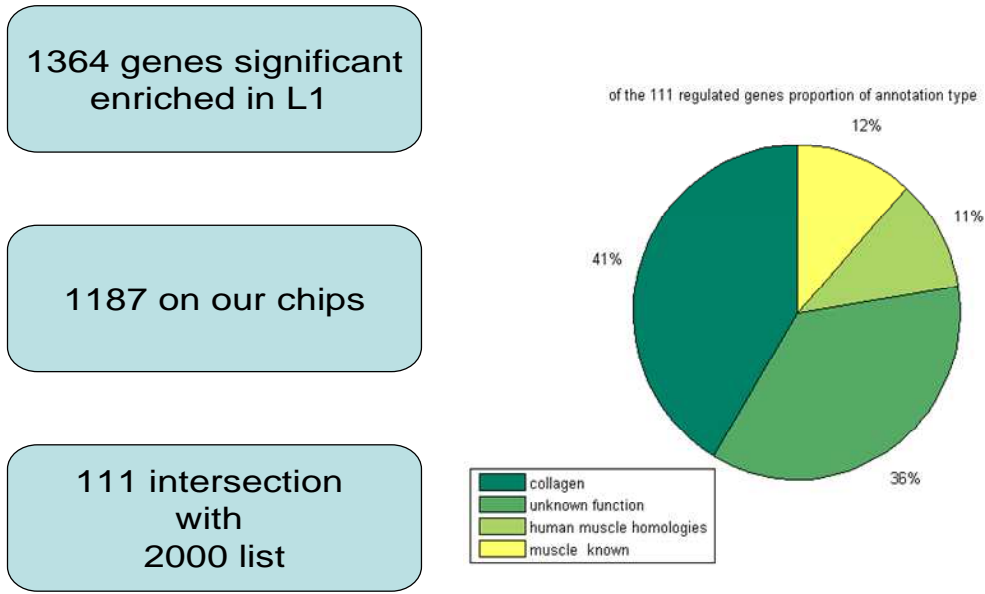

1364 genes significant enriched in L1

1187 on our chips

111 intersection with 2000 list

of the 111 regulated genes proportion of annotation type

12%
11%
41%
36%

- collagen
- unknown function
- human muscle homologies
- muscle  known

Fig. 29  Lists  of genes used in  Kim experiment analyzes.

The total genes with muscle or human homologies is 25.  See Table 10   below for  these 25 genes.

F46H5.3 Member of the arginine kinase, phosphotransferase protein family

F07A5.7 Paramyosin, major component of muscle filaments, structural equivalent of the rod region of myosin heavy chains

F53A9.10 Troponin T, putative paralog of C. elegans F53A9.10 protein; human homologus: Splice Isoform 8 of Troponin T, cardiac muscle
T22E5.5 Putative troponin-T, has strong similarity to C. elegans MUP-2 and D. melanogaster UP (upheld) troponin-T proteins; human homologus-Splice Isoform 8 of Troponin T, cardiac muscle

T25F10.6 Putative paralog of C. elegans UNC-87 which encodes muscle thin filament-associated protein

F42G4.3 ZYX-1 is potentially orthologous to the human gene LIPOMA-PREFERRED PARTNER (LPP, OMIM:600700). [detailsProtein with strong similarity to members of the LIM domain containing protein family

W05G11.6human homolog; Member of the phosphoenolpyruvate carboxykinase protein family

H22K11.1 Probable aspartyl protease and an ortholog of human cathepsin D

Y38F1A.9 Putative member of Immunoglobulin superfamily

F11C3.3 Sarcomeric Myosin Heavy Chain, major component of thick filaments in body-wall muscle

F09B9.4 Protein with weak similarity to S. cerevisiae YDL099W

T20B3.2 Putative troponin-I ; human homologus:Troponin I, cardiac muscle

H14N18.1 Highly similar to mammalian BAG-2, BCL2-associated athanogene 2, a chaperone regulator

W01F3.3/mlt-11; simmilarity with human: Splice Isoform Alpha of Tissue factor pathway inhibitor precursor

F40E10.3 Protein with strong similarity to human CASQ2 protein, a cardiac muscle calsequestrin 2

C16A3.6 Protein with strong similarity to S. cerevisiae Mak16p, an essential nuclear protein required for propagation of M1 double-stranded RNA; human homolog:RNA binding protein

C38C6.4/sre-13 G protein-coupled receptor, member of a subfamily with SRE proteins which are expressed in chemosensory neurons, no homolog found in humans...

F55B11.3 Protein with strong similarity to H. sapiens Hs.169504 gene product [Human mRNA for KIAA0170 gene (GenBank)]

F38C2.5 Zinc finger protein with strong similarity to C. elegans Y57G11C.25 and C. elegans POS-1, a cytoplasmic zinc-finger protein involved in ...

T09A5.6- Component of the Mediator complex required for transcriptional regulation of certain genes human homolog-Hypothetical protein MGC5309

ZC101.2 Muscle protein that is a member of the Immunoglobulin superfamily

C24A3.5 Member of the 4 TM potassium channel protein family

 C13C12.1
Calmodulin

F27D9.8 Protein with strong similarity to human Hs.172278 protein,
beta2-syntrophin

K07A9.2 Serine/threonine protein kinase, has similarity to human, D. melanogaster, and S.
cerevisiae calcium/calmodulin-dependent protein kinase...

 Table 10. 25 genes with muscle or human homologues.

The gene expression pattern of the 111 genes on our array is of low gene expression
for the mid life time points day 9,10,11 and high gene expression at the beginning of
the adult life day3 and the end, day 12,15 of this nematode.

**111 genes enriched in muscle with highest variation in our data:**
**During the midlife time points can be seen a**
**overall trend of decreasing of gene expression to be followed by a**
   **slight increase in gene expression.**

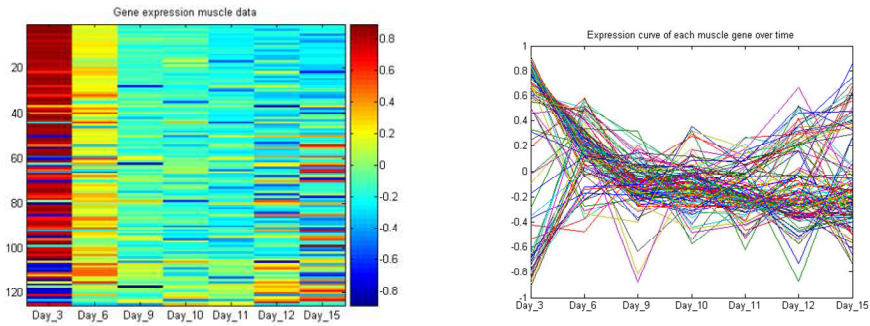

**Fig. 30 left -heat map of 111 genes, right –gene expression over the time points day3-day15**

4.4 *C. elegans* and mammalian muscle homologies

One of the final steps in my sarcopenia analysis was to compile mammalian muscle related genes and find their homologues in *C. elegans*. I will just mention here the lists of genes I compiled. Future analysis remains to be done as comparing the lists I compiled with other experiments. In this sense I will also mention David Miller's recent work on embryonic muscle transcriptome of Caenorhabditis elegans and how his list of human homologues overlaps with my lists.

Using NCBI data base and SQL language I searched for genes mammalian muscle related genes. I identified 4136 genes. Among these genes I searched the genes that have homologies in general and identified 1488 genes. Next step was to look for genes that have *C. elegans* homologues among the 1488 genes. I identified 325 genes. In this list of 325 genes 21 are muscle related genes, 10 aging related genes and the rest are other genes related with various other processes as: cyclin dependent kinase family, abnormal chemotaxis, abnormal cell linage, defective laying eggs, kinase proteins.

See Appendix C for the Table with 325 genes.

Another search I've done was for identifying mammalian muscle genes involved in senescence. I compiled a list of 43 genes. In this list 7 genes have C. elegans homologies, see below.

Below are the 7 genes

| **1: laminin** |
| --- |
| **K08C7.3a [**Caenorhabditis elegans]Other Aliases: K08C7.3Other Designations: |
| K08C7.3bChromosome: IV |
| **2: *sir-2.1*** |
| yeast SIR related [Caenorhabditis elegans]Other Aliases: R11A8.4Other |
| Designations: yeast SIR related family member (sir-2.1)Chromosome: IV |
| **3: *pab*-3** |
| PolyA Binding protein [Caenorhabditis elegans]Other Aliases: C17E4.5Other |
| Designations: PolyA Binding protein family member (pab-3)Chromosome: |
| **4: *hlh*-2** |
| Helix Loop Helix [Caenorhabditis elegans]Other Aliases: M05B5.5Other |
| Designations: Helix Loop Helix family member (hlh-2)Chromosome: |
| **5: *pmk*-1** |
| P38 Map Kinase family [Caenorhabditis elegans]Other Aliases: B0218.3Other |
| Designations: P38 Map Kinase family member (pmk-1)Chromosome: IV |
| **6: *ced*-10** |
| CEll Death abnormality [Caenorhabditis elegans]Other Aliases: C09G12.8Other |
| Designations: CEll Death abnormality family member (ced-10)Chromosome: IV |
| **7: *mpk*-1** |

| MAP Kinase [Caenorhabditis elegans]Other Aliases: F43C1.2Other Designations: MAP |
| Kinase family member (mpk-1)Chromosome: III |

 I also compiled  a list of  119  mammalian muscle-related genes known to be involved in aging. In this list 20 genes have *C. elegans* homologies. 16 out of these 20 genes are on our chips. The graphs with the 16 genes expression  as well as a Table with  the 20 genes can be seen in Appendix D.  Just four genes out of these 20 genes are expressed in cell muscle and none of them are among our list with 2000 genes with highest variation in our experiment.

The four genes are:

'F10C1.2' ifb-1
'C12D8.10' akt-1
'ZK792.6' let-60
'C29F9.7'  pat-4

David Miller group performed also relatively recent (2007) an experiment with the purpose of analyzing the embryonic muscle transcriptome of Caenorhabditis elegans. They have applied Micro-Array Profiling of Caenorhabditis elegans Cells (MAPCeL) to muscle cell populations extracted from developing Caenorhabditis elegans embryos. Fluorescence Activated Cell Sorting (FACS) was used to isolate myo-3::GFP-positive muscle cells, and their cultured derivatives, from dissociated early Caenorhabditis elegans embryos. Microarray analysis identified 6,693 expressed genes, 1,305 of which are enriched in the myo-3::GFP positive cell population relative to the average embryonic cell. The muscle-enriched gene set was validated by comparisons to known muscle markers, independently derived expression data, and GFP reporters in transgenic strains. This study provides a comprehensive description of gene expression in developing Caenorhabditis elegans embryonic muscle cells. They founded that over half of the muscle-enriched transcripts encode proteins with human homologs suggesting that mutant analysis of these genes in Caenorhabditis elegans could reveal evolutionarily conserved models of muscle gene function with ready application to human muscle pathologies.

I used David Miller's human homologies genes list of 788 genes suggested in this study to compare with my lists. Out of these 788 genes, 593 are in our experiment and all are in the list of 721 genes I compiled. Just as a reminder, the 721 gene list are that genes I identified to be expressed in muscle cell using AQL language.

Also I checked for the intersection of the 593 genes in the list of 2000 genes from our experiment with highest variation and identified 61 genes. A majority of these genes can be found in the cluster G18 mentioned in Chapter 1.

## 5. Summarizing and conclusions

Using various bioinformatics tools I compiled various gene lists muscle-related. See Fig. 31 below. When cluster the 42 genes that represents genes expressed in cell muscle with high variation in our experiment we identify more the 50% of this genes as having the expected sarcopenia signature of down-regulated genes pattern. I called such pattern and the genes with such pattern as having a 'positive connection' with the sarcopenia phenotype. All this genes are in the cluster G6 (see section 4.1). The main biological theme for these genes is involvement in growth, developmental and defective in locomotion. In the same time we identified an up-regulated pattern in the cluster G7 (again section 4.1). The genes in this cluster might suggest further bench experiments as I will discuss below.

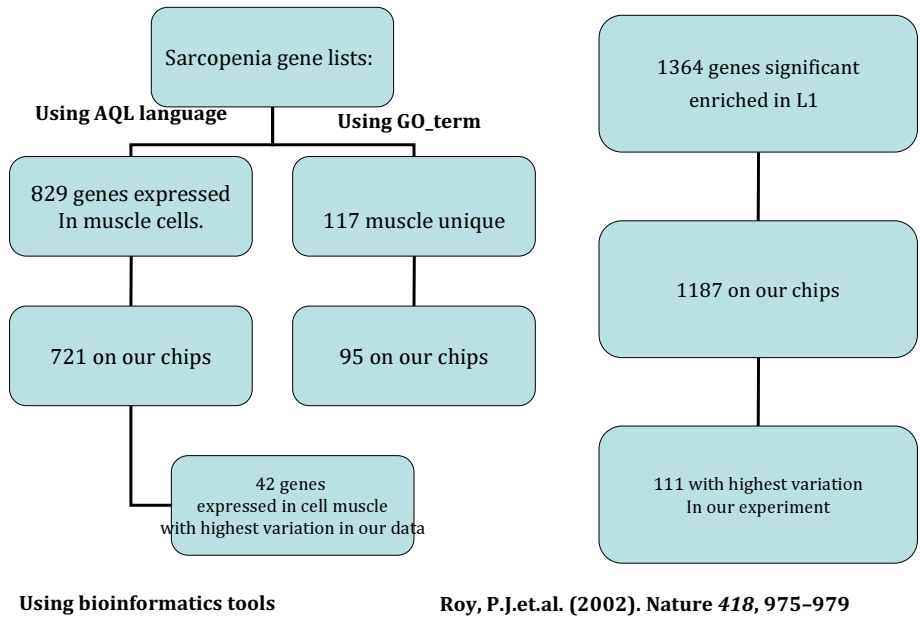

Fig. 31 gene lists used in Sarcopenia analysis

For example, one might want to know what might happen if the genes in cluster G7 introduced in section 4.1 would have a down-regulated pattern instead of an increase in gene expression as these genes show in our analyzes of the wild type of this nematode. One might want to understand if an reverse in gene expression output for the genes in cluster G7 might slow down the sarcopenia signs.

When I analyzed the 95 gene list using the GO-term I identified 2 main patterns of up-regulated and down-regulated genes. The genes in the down-regulated category show same 'positive connection' with the signature of sarcopenia as identified in section 4.1 This genes are mainly structural genes and found in cluster G8 (see section 4.2 and Appendix B).

Apart of these findings I also analyzed a list of genes significantly enriched in muscle cells of young animal (L1 stage). This list was obtained from an experiment performed by Kim's group. I identified 111 genes to have a high variation in our experiment. These genes are mainly collagen related genes. 25 genes have muscle related function, see Table 10.

*C. elegans* body wall muscle undergoes a process remarkably reminiscent of human sarcopenia. Both have mid-life onset and are characterized by progressive loss of sarcomeres and cytoplasmic volume; both are associated with locomotory decline. To extend understanding of this fundamental problem, I have focused microarray analyses on *C.elegans* muscle aging. Genes expressed in muscle as well as muscle related have been identified and emergent patterns in this list were defined. I surveyed expression of all to describe a profile of transcriptional changes in muscle that transpires during adult life and aging.

This research describes age-associated changes in muscle gene transcription and will constitute the first full-genome profile of sarcopenia in any animal.

## Acknowledgements

This work would have not been possible without the constant discussions with Prof. Monica Driscoll. The microarray raw data are of Driscoll lab. Also I have to acknowledge Prof. Eytan Domany who helped me understand his clustering method during a stage at Weizmann Institute, Institute of Complex Systems. Nevertheless I have to acknowledge Beate Hartmann., Christophe Grundschober, and Patrick Nef from Hoffmann-LaRoche, Basel, Switzerland, who helped providing the raw data to Driscoll Lab. Most of this work was supported by a grant of Prof Monica Driscoll.

**Appendix A**

**Biological process: muscle development:**

**Definition:** The process whose specific outcome is the progression of the muscle over time, from its formation to the mature structure.

| act-1 | T04C12.6 | An actin that affects body wall and pharyngeal muscle |
|-------|----------|-------------------------------------------------------|
| eat-1 | T11B7.4 | eat-1 encodes both a homolog of mammalian alpha-ac |
| hlh-1 | B0304.1 | hlh-1 encodes a basic helix-loop-helix (bHLH) |
| mlc-1 | C36E6.3 | encodes a muscle regulatory myosin light chain tha |
| syd-2 | F59F5.6 | syd-2 encodes alpha-liprin, a member of the liprin |
| tmd-2 | C08D8.2 | |
| unc-52 | ZC101.2 | The unc-52 gene encodes perlecan, a protein orthol |
| unc-89 | C09D1.1 | |

**Biological process: pharyngeal muscle development**

Definition: The process whose specific outcome is the progression of the pharyngeal muscle over time, from its formation to the mature structure.

| eff-1 | C26D10.5 | The eff-1 gene encodes a novel, type I transmembra |
|-------|----------|---------------------------------------------------|
| glp-1 | F02A9.6 | glp-1 encodes an N-glycosylated transmembrane prot |
| pop-1 | W10C8.2 | pop-1 encodes an HMG box-containing protein that i |

## Biological process: muscle cell fate specification

**Definition:** Process by which a cell becomes capable of differentiating autonomously into a muscle cell in an environment that is neutral with respect to the developmental pathway; upon specification, the cell fate can be reversed.

| mls-1 | H14A12.4 | mls-1 encodes a T-box transcription factor |
|-------|----------|---------------------------------------------|

## Molecular function: structural constituent of muscle

**Definition:** The action of a molecule that contributes to the structural integrity of a muscle fiber.

| pat-10 | F54C1.7 | pat-10 encodes body wall muscle troponin C, |
|--------|---------|---------------------------------------------|
| tmd-2 | C08D8.2 | |
| unc-89 | C09D1.1 | |

## Major muscle structural genes:

**myosin:**

**biological process**: **muscle contraction**: A process leading to shortening and/or development of tension in muscle tissue. Muscle contraction occurs by a sliding filament mechanism whereby actin filaments slide inward among the myosin filaments.

| egl-2 | F16B3.1 | egl-2 encodes a voltage-gated potassium channel |
|---|---|---|
| itr-1 | F33D4.2 | itr-1 encodes a putative IP3 receptor |
| jph-1 | T22C1.7 | jph-1 encodes a junctophilin, |
| pat-10 | F54C1.7 | pat-10 encodes body wall muscle troponin C |
| twk-18 | C24A3.6 | twk-18 encodes one of 44 C. elegans TWK |
| unc-26 | JC8.10 | unc-26 encodes synaptojanin, a polyphosphoinositid |

Cellular-component: actomyosin, actin component

Definition: The actin part of any complex of actin, myosin, and accessory proteins.

| unc-78 | C04F6.4 | The unc-78 gene encodes a homolog of actin |
|---|---|---|

**Cellular-component**: **contractile fiber**

**Definition:** Fibers, composed of actin, myosin, and associated proteins, found in cells of smooth or striated muscle.

| pat-10 | F54C1.7 | pat-10 encodes body wall muscle troponin C |
|---|---|---|

**Cellular-component: muscle myosin**

**Definition:** A filament of myosin found in a muscle cell of any type.

| mlc-1 | C36E6.3 | encodes a muscle regulatory myosin |
|---|---|---|

**Cellular-component: myosin**

**Definition:** A protein complex, formed of one or more myosin heavy chains plus associated light chains and other proteins, that functions as a molecular motor; uses the energy of ATP hydrolysis to move actin filaments or to move vesicles or other cargo on fixed actin filaments; has magnesium-ATPase activity and binds actin. Myosin classes are distinguished based on sequence features of the motor, or head, domain, but also have distinct tail regions that are believed to bind specific cargoes.

| hum-1 | F29D10.4 | |
|---|---|---|
| hum-2 | F36D4.3 | hum-2 is orthologous to the human gene MYOSIN HEAV |
| hum-4 | F46C3.3 | hum-4 encodes a class XII myosin, |
| hum-5 | T02C12.1 | |
| hum-6 | T10H10.1 | hum-6 is orthologous to the human gene MYOSIN VIIA |
| hum-7 | F56A6.2 | hum-7 encodes a class IX unconventional myosin |
| hum-8 | Y66H1A.6 | |
| let-75 | R06C7.10 | let-75 encodes a pharynx-specific type II myosin |
| myo-2 | T18D3.4 | myo-2 encodes a muscle-type specific myosin heavy |
| myo-3 | K12F2.1 | myo-3 encodes MHC A, the minor isoform of MHC |
| nmy-1 | F52B10.1 | nmy-1 encodes a class II non-muscle myosin heavy chain |
| nmy-2 | F20G4.3 | nmy-2 encodes a maternally expressed nonmuscle myo |

| spe-15 | F47G6.4 | spe-15 is orthologous to the human gene MYOSIN VI |
|--------|---------|---------------------------------------------------|
| unc-15 | F07A5.7 | The unc-15 gene encodes a paramyosin ortholog; UNC |
| unc-54 | F11C3.3 | unc-54 encodes a muscle myosin class II heavy chain |
|        | F43C9.3 | |
|        | F45G2.2 | |
|        | F58G4.1 | |
|        | W03F11.6 | |
|        | Y11D7A.14 | |

**Cellular-component**: **myosin II**

**Definition:** A myosin complex containing two class II myosin heavy chains, two myosin essential light chains and two myosin regulatory light chains. Also known as classical myosin or conventional myosin, the myosin II class includes the major muscle myosin of vertebrate and invertebrate muscle, and is characterized by alpha-helical coiled coil tails that self assemble to form a variety of filament structures.

| myo-2 | T18D3.4 | myo-2 encodes a muscle-type specific myosin heavy |
|-------|---------|---------------------------------------------------|
| nmy-2 | F20G4.3 | nmy-2 encodes a maternally expressed nonmuscle myo |

**Actin:**

**Biological function:** actin cytoskeleton organization and biogenesis

**Definition:** The assembly and arrangement of cytoskeletal structures comprising actin filaments and their associated proteins.

| act-1 | T04C12.6 | An actin that affects body wall and pharyngeal mus |
|-------|----------|---------------------------------------------------|
| act-4 | M03F4.2 | An actin that is expressed in body wall and vulval |
| cap-1 | D2024.6 | cap-1 encodes an F-actin capping protein alpha sub |
| cap-2 | M106.5 | The beta subunit of actin capping protein that reg |
| cyk-1 | F11H8.4 | The cyk-1 gene encodes a homolog of Drosophila |
| fhod-1 | C46H11.11 | |
| fhod-2 | F56E10.2 | |
| fozi-1 | K01B6.1 | K01B6.1 encodes a protein with a zinc-finger domain |
| pfn-1 | Y18D10A.20 | |
| pfn-3 | K03E6.6 | |
| tag-268 | F58B6.2 | |
| unc-53 | F45E10.1 | UNC-53 encodes at least five large (~1200-1600 residues |
| | F15B9.4 | |
| | F56E10.3 | |
| | Y48G9A.4 | |

Biological function: muscle contraction

Definition: A process leading to shortening and/or development of tension in muscle tissue. Muscle contraction occurs by a sliding filament mechanism whereby actin filaments slide inward among the myosin filaments.

| | | |
|---|---|---|
| egl-2 | F16B3.1 | egl-2 encodes a voltage-gated potassium channel |
| itr-1 | F33D4.2 | itr-1 encodes a putative inositol (1,4,5) |
| jph-1 | T22C1.7 | jph-1 encodes a junctophilin, |
| pat-10 | F54C1.7 | pat-10 encodes body wall muscle troponin C, |
| twk-18 | C24A3.6 | |
| unc-26 | JC8.10 | unc-26 encodes synaptojanin, a polyphosphoinositid |

**Biological function:actin filament organization**

**Definition:** Control of the spatial distribution of actin filaments; includes organizing filaments into meshworks, bundles, or other structures, as by cross-linking.

| ced-12 | Y106G6E.5 | ced-12 is required both for phagocytotic engulfment |
|--------|-----------|-----------------------------------------------------|
| die-1  | C18D1.1   | die-1 encodes a C2H2 zinc finger protein            |
| wve-1  | R06C1.3   | wve-1 encodes a homolog of the mammalian WAVE protein |

## <u>Biological function:</u>**barbed-end actin filament capping**

Definition: The binding of a protein or protein complex to the barbed (or plus) end of an actin filament, thus preventing the addition or exchange of further actin subunits. The pointed (or minus) ends of fragments remain uncapped and are rapidly shortened.

| add-1  | F39C12.2 | add-1 encodes an ortholog of the cytoskeletal prot |
|--------|----------|----------------------------------------------------|
| unc-78 | C04F6.4  |                                                    |

<u>Biological function:</u> cortical actin cytoskeleton organization and biogenesis

Definition: The assembly and arrangement of actin-based cytoskeletal structures in the cell cortex, i.e. just beneath the plasma membrane.

| add-1 | F39C12.2 | add-1 encodes an ortholog of the cytoskeletal prot |
|-------|----------|----------------------------------------------------|
| add-2 | F57F5.4  | add-2 encodes a protein containing an alpha-adduci |

**Cellular component: adherens junction**

**Definition:** A cell junction at which the cytoplasmic face of the plasma membrane is attached to actin filaments.

| dlg-1 | C25F6.2 | dlg-1 encodes a MAGUK protein, orthologous to Drosophila |
|---|---|---|
| pat-4 | C29F9.7 | The pat-4 gene encodes a serine/threonine kinase |
| ptp-3 | C09D8.1 | ptp-3 encodes a receptor-like tyrosine phosphatase |
| unc-97 | F14D12.2 | The unc-97 gene encodes a LIM domain |

**Cellular component:cortical actin cytoskeleton**

**Definition:** The portion of the actin cytoskeleton that lies just beneath the plasma membrane.

| add-1 | F39C12.2 | ortholog of the cytoskeletal protein |
|---|---|---|
| add-2 | F57F5.4 | |

Cellular component:dynactin complex

Definition: A 20S multiprotein assembly of total mass about 1.2 MDa that activates dynein-based activity in vivo. A large structural component of the complex is an actin-like 40 nm filament composed of actin-related protein, to which other components attach.

| dnc-1 | ZK593.5 | DNC-1 encodes a dynactin, orthologous to Drosophil |
|---|---|---|
| dnc-2 | C28H8.12 | dnc-2 encodes a member of the dynamitin family |

## Molecular function: actin binding

**Definition:** Interacting selectively with monomeric or multimeric forms of actin, including actin filaments.

| | | |
|---|---|---|
| add-1 | F39C12.2 | add-1 encodes an ortholog of the cytoskeletal prot |
| anc-1 | ZK973.6 | anc-1 encodes a 8546-residue protein, orthologous |
| atn-1 | W04D2.1 | atn-1 encodes an alpha-actinin homolog. |
| cap-1 | D2024.6 | cap-1 encodes an F-actin capping protein alpha sub |
| cap-2 | M106.5 | The beta subunit of actin capping protein that reg |
| cyk-1 | F11H8.4 | The cyk-1 gene encodes a homolog of Drosophila dia |
| ers-2 | ZC434.5 | ers-2 encodes a predicted mitochondrial glutamyl-t |
| fhod-1 | C46H11.11 | |
| fhod-2 | F56E10.2 | |
| fli-1 | B0523.5 | fli-1 encodes a homolog of Drosophila flightless-I |
| fozi-1 | K01B6.1 | K01B6.1 encodes a protein with a zinc-finger domai |
| frm-1 | ZK270.2 | |
| pfn-1 | Y18D10A.20 | |
| pfn-3 | K03E6.6 | |
| sma-1 | R31.1 | |
| tag-268 | F58B6.2 | |
| tth-1 | F08F1.8 | |
| unc-115 | F09B9.2 | unc-115 encodes a protein that binds actin filamen |
| unc-60 | C38C3.5 | unc-60 encodes orthologs of actin depolymerizing f |
| unc-70 | K11C4.3 | unc-70 encodes two isoforms of a beta-spectrin ort |
| unc-78 | C04F6.4 | The unc-78 gene encodes a homolog of actin-interac |
| vab-10 | ZK1151.1 | vab-10 encodes, by alternative splicing, two spect |
| | C10H11.1 | |
| | F15B9.4 | |
| | F38E9.5 | |
| | F56E10.3 | |
| | K06A4.3 | |
| | K08E3.4 | |
| | M116.5 | |
| | Y48G9A.4 | |
| | Y50D7A.10 | |
| | Y71G12B.11 | |
| | ZK370.3 | |

**Molecular function** **calmodulin binding**

**Definition:** Interacting selectively with calmodulin, a calcium-binding protein with many roles, both in the calcium-bound and calcium-free states.

| | | |
|---|---|---|
| add-1 | F39C12.2 | add-1 encodes an ortholog of the cytoskeletal |
| hda-4 | C10E2.3 | hda-4 encodes a class II histone deacetylase |
| | B0399.1 | |
| | C03F11.1 | |
| | C14B9.8 | C14B9.8 is orthologous to the human gene PHOSPHORYLASE KINASE alpha 2 |
| | C53A5.5 | |
| | F08A10.1 | |
| | Y50D7A.3 | Y50D7A.3 is orthologous to human PHOSPHORYLASE KINASE gamma 2 |
| | Y67D8A.1 | |
| | Y67D8A.2 | |

**Molecular  function:** **cytoskeletal protein binding**

**Definition:** Interacting selectively with any protein component of any cytoskeleton (actin, microtubule, or intermediate filament cytoskeleton).

| | | |
|---|---|---|
| ajm-1 | C25A11.4 | ajm-1 encodes a member of the apical junction molecule |
| dnc-1 | ZK593.5 | DNC-1 encodes a dynactin, orthologous to Drosophila |
| erm-1 | C01G8.5 | The erm-1 gene encodes an ortholog of the ERM family |
| frm-1 | ZK270.2 | |
| frm-2 | T04C9.6 | |
| frm-3 | H05G16.1 | frm-3 encodes a protein containing a FERM domain |
| nfm-1 | F42A10.2 | nfm-1 encodes a homolog of human merlin |
| ptp-1 | C48D5.2 | ptp-1 encodes a non-receptor tyrosine phosphatase |
| sdn-1 | F57C7.3 | sdn-1 encodes a homolog of vertebrate syndecan-2, |

**Molecular_function microtubule binding**

**Definition:** Interacting selectively with microtubules, filaments composed of tubulin monomers.

| | | |
|---|---|---|
| dnc-1 | ZK593.5 | DNC-1 encodes a dynactin, orthologous to Drosophila |
| ebp-1 | Y59A8B.7 | |
| ebp-2 | VW02B12L.3 | |
| osm-3 | M02B7.3 | osm-3 encodes a homolog of the heavy chain subunit |
| tag-201 | Y59A8B.9 | |
| unc-104 | C52E12.2 | The unc-104 gene encodes a kinesin-like motor protein |
| zyg-8 | Y79H2A.11 | The zyg-8 gene encodes a protein with a doublecort |
| zyg-9 | F22B5.7 | |

**Molecular function: protein binding**

**Definition:** Interacting selectively with any protein or protein complex (a complex of two or more proteins that may include other non protein molecules).

| | | |
|---|---|---|
| act-1 | T04C12.6 | An actin that affects body wall and pharyngeal muscle |
| act-2 | T04C12.5 | act-2 encodes an invertebrate actin |
| act-3 | T04C12.4 | act-3 encodes an invertebrate actin, |
| act-4 | M03F4.2 | An actin that is expressed in body wall |
| act-5 | T25C8.2 | an ortholog of human cytoplasmic actin; |
| apc-10 | F15H10.3 | apc-10 encodes a homolog of the fission yeast |
| aph-1 | VF36H2L.1 | |
| arx-1 | Y71F9AL.16 | |
| arx-2 | K07C5.1 | |
| athp-1 | C44B9.4 | |

**Molecular function spectrin binding**

**Definition:** Interacting selectively with spectrin, a protein that is the major constituent of the erythrocyte cytoskeletal network. It associates with band 4.1 (see band protein) and actin to form the cytoskeletal superstructure of the erythrocyte plasma membrane. It is composed of nonhomologous chains, alpha and beta, which aggregate side-to-side in an antiparallel fashion to form dimers, tetramers, and higher polymers.

| add-1 | F39C12.2 | add-1 encodes an ortholog of the cytoskeletal prot |
|-------|----------|----------------------------------------------------|

**Cellular component(the protein is localized within the cell**: presynaptic membrane

**Definition:** A specialized area of membrane of the axon terminal that faces the plasma membrane of the neuron or muscle fiber with which the axon terminal establishes a synaptic junction; many synaptic junctions exhibit structural presynaptic characteristics, such as conical, electron-dense internal protrusions, that distinguish it from the remainder of the axon plasma membrane.

| egl-47 | C50H2.2 | The egl-47 gene encodes a G protein-coupled receptor |
|--------|---------|------------------------------------------------------|
| unc-11 | C32E8.10 | |

**Troponin:**

**troponin complex Type: Cellular_component**

**Definition:** A complex of accessory proteins (typically troponin T, troponin I and troponin C) found associated with actin in muscle thin filaments; involved in calcium regulation of muscle contraction.

| mup-2 | T22E5.5 | mup-2 encodes the muscle contractile protein tropo |
|---|---|---|

**tropomyosin binding** Type: Molecular_function

**Definition:** Interacting selectively with tropomyosin, a protein associated with actin filaments both in cytoplasm and, in association with troponin, in the thin filament of striated muscle.

| tmd-1 | C06A5.7 | tmd-1 encodes a predicted member of the tropomodul |
|---|---|---|
| tmd-2 | C08D8.2 | |

**Appendix D**

The 20 genes  homologues genes identified in a list of **119 mammalian muscle gene aging related. 4 of this 20 genes are found in worm data base to be expressed in cell muscle: they are:**
'F10C1.2' ifb-1
'C12D8.10' akt-1
'ZK792.6' let-60
'C29F9.7'  pat-4

16 genes out of the 20 are on our chips. 4 graphs representing their expression over *C.Elegans* life span. On the axis Y is the absolute values of the gene expressions in log2 scale. On X axis are the 7 time points.

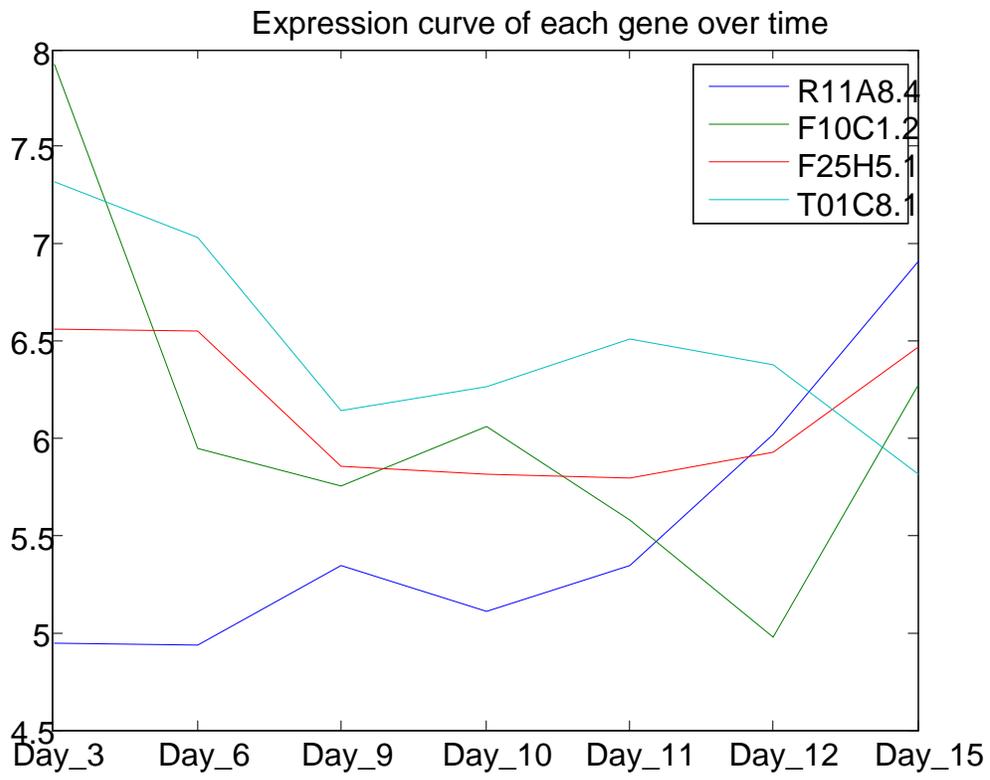

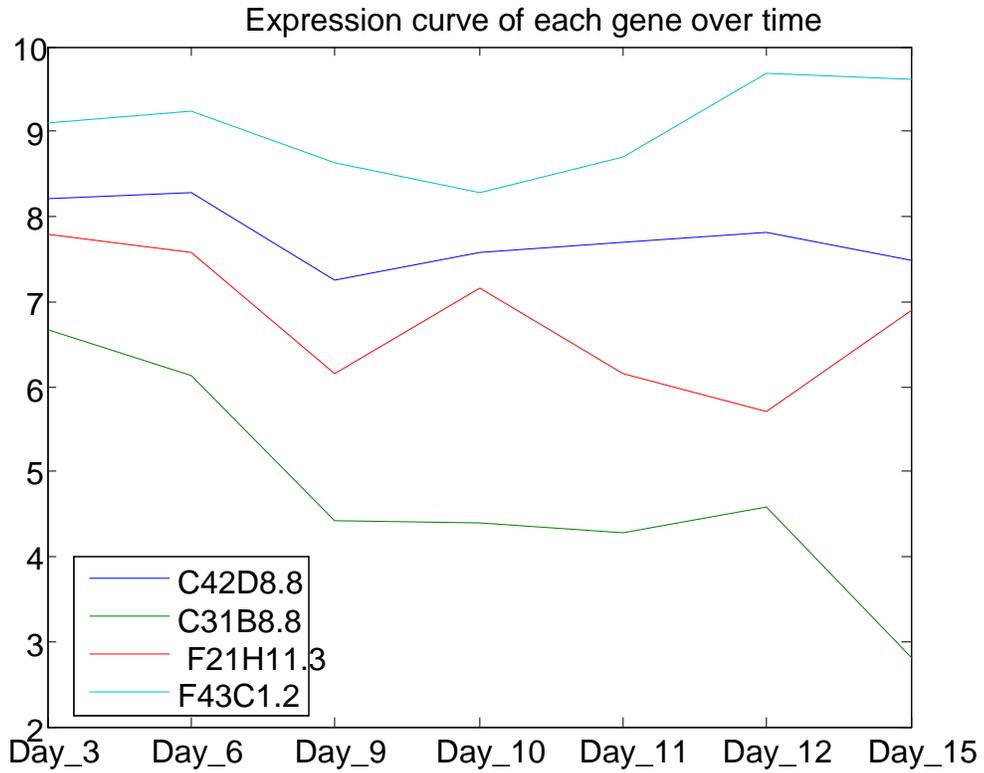

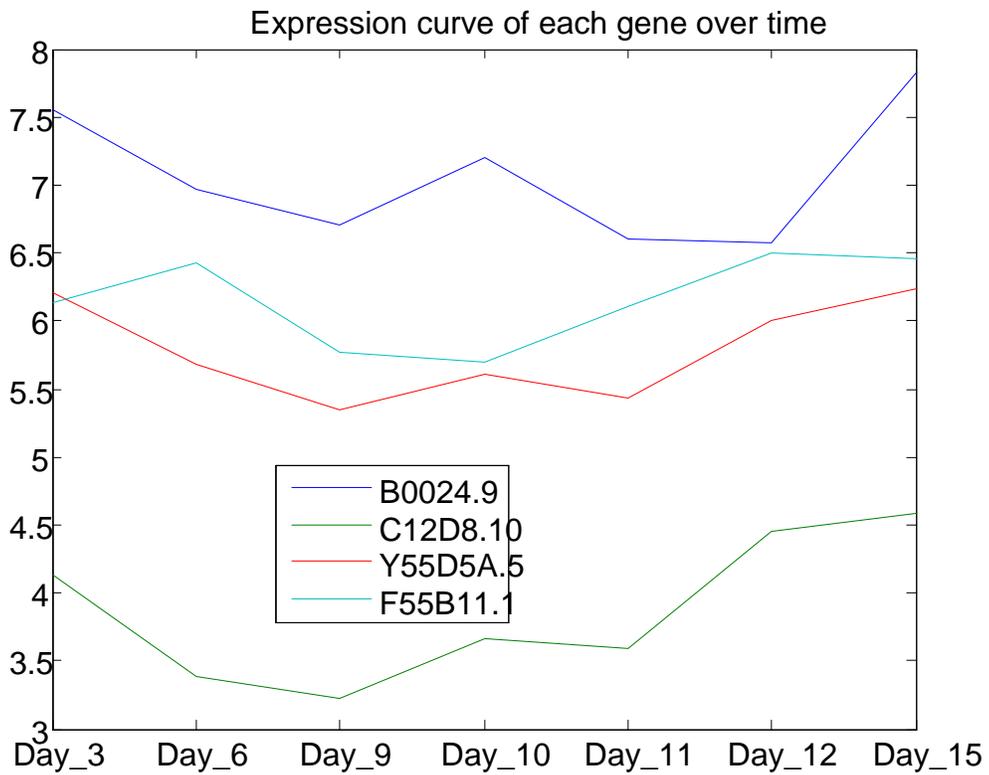

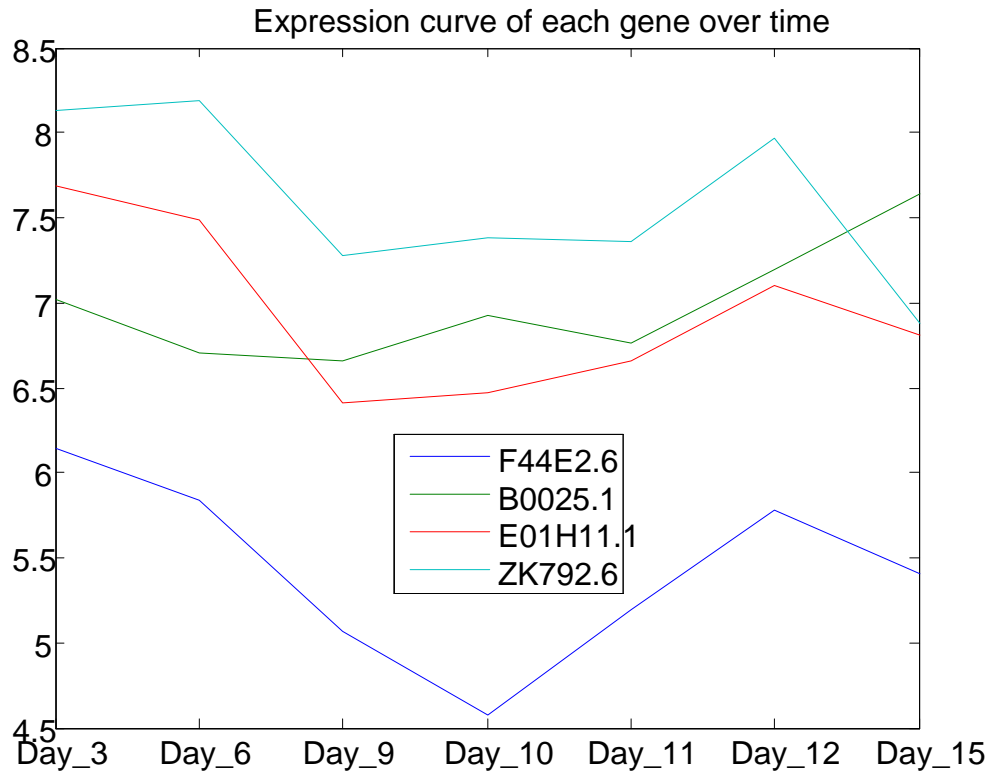

**Mammalian muscle aging homologues in C.Elegans:**

| |
|---|
| **1: sir-2.1** |
| yeast SIR related [Caenorhabditis elegans]Other Aliases: R11A8.4Other Designations: yeast SIR related family member (sir-2.1)Chromosome: IV Annotation: NC_003282.4 (10366208..10363296, complement)GeneID: 177924 |
| **2: ifb-1** |
| Intermediate Filament, B [Caenorhabditis elegans]Other Aliases: F10C1.2Other Designations: Intermediate Filament, B family member (ifb-1)Chromosome: II Annotation: NC_003280.6 (5765583..5772103)GeneID: 173976 |
| **3: tag-15** |
| Temporarily Assigned Gene name [Caenorhabditis elegans]Other Aliases: F25H5.1Other Designations: Temporarily Assigned Gene name family member (tag-15)Chromosome: I Annotation: NC_003279.5 (9210329..9181809, complement)GeneID: 3565712 |
| **4: COX1** |
| cytochrome c oxidase subunit I [Caenorhabditis elegans]Mitochondrion: MTAnnotation: NC_001328.1 (7844..9421)GeneID: 2565700 |
| **5: aak-2** |

AMP-Activated Kinase [Caenorhabditis elegans]Other Aliases: T01C8.1Other Designations: AMP-Activated Kinase family member (aak-2)Chromosome: X Annotation: NC_003284.6 (16796001..16803271)GeneID: 181727

**6: apl-1**

Amyloid Precursor-Like [Caenorhabditis elegans]Other Aliases: C42D8.8Other Designations: Amyloid Precursor-Like family member (apl-1)Chromosome: X Annotation: NC_003284.6 (5117308..5112388, complement)GeneID: 180783

**7: C31B8.8**

C31B8.8 [Caenorhabditis elegans]Other Aliases: C31B8.8Chromosome: V Annotation: NC_003283.7 (2909509..2905596, complement)GeneID: 178748

**8: tbx-2**

T BoX family [Caenorhabditis elegans]Other Aliases: F21H11.3Other Designations: T BoX family member (tbx-2)Chromosome: III Annotation: NC_003281.7 (5136277..5130885, complement)GeneID: 175698

**9: mpk-1**

MAP Kinase [Caenorhabditis elegans]Other Aliases: F43C1.2Other Designations: MAP Kinase family member (mpk-1)Chromosome: III Annotation: NC_003281.7 (4228065..4216735, complement)GeneID: 175545

**10:** thioredoxin

**B0024.9** [Caenorhabditis elegans]Other Aliases: B0024.9Chromosome: V Annotation: NC_003283.7 (10314438..10315628)GeneID: 179434

**11: akt-1**

AKT kinase family [Caenorhabditis elegans]Other Aliases: C12D8.1Other Designations: AKT kinase family member (akt-1)Chromosome: V Annotation: NC_003283.7 (10248703..10253259)GeneID: 179424

**12: daf-2**

abnormal Dauer Formation [Caenorhabditis elegans]Other Aliases: Y55D5A.5Other Designations: abnormal DAuer Formation family member (daf-2)Chromosome: III Annotation: NC_003281.7 (3028789..2995752, complement)GeneID: 175410

13: F55B11.1

**F55B11.1** [Caenorhabditis elegans]Other Aliases: F55B11.1Chromosome: IV Annotation: NC_003282.4 (14412851..14401100, complement)GeneID: 178381

14: F44E2.6

**F44E2.6b** [Caenorhabditis elegans]Other Aliases: F44E2.6Other Designations: F44E2.6aChromosome: III Annotation: NC_003281.7 (8845091..8844478, complement)GeneID: 176244

**15: vps-34**

related to yeast Vacuolar Protein Sorting factor [Caenorhabditis elegans]Other Aliases: B0025.1Other Designations: related to yeast Vacuolar Protein Sorting factor family member (vps-34)Chromosome: I Annotation: NC_003279.5 (6028487..6033458)GeneID: 172280

| |
|---|
| **16: pkc-2** |
| Protein Kinase C [Caenorhabditis elegans]Other Aliases: E01H11.1Other Designations: Protein Kinase C family member (pkc-2)Chromosome: X Annotation: NC_003284.6 (9371060..9386328)GeneID: 181166 |
| **17: let-60** |
| LEThal [Caenorhabditis elegans]Other Aliases: ZK792.6Other Designations: LEThal family member (let-60)Chromosome: IV Annotation: NC_003282.4 (11691057..11688196, complement)GeneID: 178104 |
| **18: pat-4** |
| Paralysed Arrest at Two-fold [Caenorhabditis elegans]Other Aliases: C29F9.7Other Designations: Paralysed Arrest at Two-fold family member (pat-4)Chromosome: III Annotation: NC_003281.7 (94807..89016, complement)GeneID: 175175 |
| **19: sod-1** |
| SOD (superoxide dismutase) [Caenorhabditis elegans]Other Aliases: C15F1.7Other Designations: SOD (superoxide dismutase) family member (sod-1)Chromosome: II Annotation: NC_003280.6 (6973824..6972644, complement)GeneID: 174141 |
| **20: Catalase** |
| Y54G11A.5 [Caenorhabditis elegans]Other Aliases: Y54G11A.5Chromosome: II Annotation: NC_003280.6 (14301944..14298706, complement)GeneID: 175085 |